\documentclass[6pt]{article}
\usepackage[margin=0.8in]{geometry}
\usepackage[utf8]{inputenc}

\usepackage{setspace}

\usepackage{amsthm}
\usepackage{amsmath}
\usepackage{amssymb}
\newcommand{\indep}{\rotatebox[origin=c]{90}{$\models$}}

\usepackage{graphicx}  
\usepackage{subfig} 
\usepackage{float}  
\usepackage{array} 
\usepackage{multirow} 
\usepackage{booktabs} 
\usepackage{rotating} 

\usepackage{tikz}  

\usepackage{colortbl}

\usepackage{algorithm} 
\usepackage{algpseudocode} 

\usepackage{hyperref}

\DeclareMathOperator*{\plim}{plim}
\DeclareMathOperator*{\argmax}{arg\,max}
\DeclareMathOperator*{\argmin}{arg\,min}

\usepackage{multibib}
\newcites{main}{References}
\newcites{appendix}{Appendix References}

\title{Estimating time-varying exposure effects through continuous-time modelling in Mendelian randomization}
\author{Haodong Tian\textsuperscript{1}\thanks{haodong.tian@mrc-bsu.cam.ac.uk}, Ashish Patel\textsuperscript{1}\thanks{ashish.patel@mrc-bsu.cam.ac.uk}, Stephen Burgess\textsuperscript{1,2}\thanks{sb452@medschl.cam.ac.uk} \vspace{4mm} \\
	\textsuperscript{1} MRC Biostatistics Unit, School of Clinical Medicine,\\University of Cambridge, Cambridge, UK \vspace{2mm} \\
	\textsuperscript{2} British Heart Foundation Cardiovascular Epidemiology Unit, \\ Department of Public Health and Primary Care, \\ University of Cambridge, Cambridge, UK \vspace{2mm} }
\date{ }

\begin{document}

\maketitle

\begin{abstract}
	Mendelian randomization is an instrumental variable method that utilizes genetic information to investigate the causal effect of a modifiable exposure on an outcome. In most cases, the exposure changes over time. Understanding the time-varying causal effect of the exposure can yield detailed insights into mechanistic effects and the potential impact of public health interventions. Recently, a growing number of Mendelian randomization studies have attempted to explore time-varying causal effects. However, the proposed approaches oversimplify temporal information and rely on overly restrictive structural assumptions, limiting their reliability in addressing time-varying causal problems. This paper considers a novel approach to estimate time-varying effects through continuous-time modelling by combining functional principal component analysis and weak-instrument-robust techniques. Our method effectively utilizes available data without making strong structural assumptions and can be applied in general settings where the exposure measurements occur at different timepoints for different individuals. We demonstrate through simulations that our proposed method performs well in estimating time-varying effects and provides reliable inference results when the time-varying effect form is correctly specified. The method could theoretically be used to estimate arbitrarily complex time-varying effects. However, there is a trade-off between model complexity and instrument strength. Estimating complex time-varying effects requires instruments that are unrealistically strong. We illustrate the application of this method in a case study examining the time-varying effects of systolic blood pressure on urea levels.
	
\end{abstract}

\newpage
\doublespacing
\section{Introduction}
Mendelian randomization (MR) utilizes genetic variants as instrumental variables (IVs) to assess the causal effects of modifiable exposures on health outcomes \cite{davey2003mendelian,lawlor2008mendelian}. In reality, most modifiable exposures are subject to change over time, indicating that the exposure effects on outcomes should be modelled over time to incorporate time-varying patterns. Examining time-varying effects can provide valuable insights into underlying mechanisms. For example, such analyses can provide evidence on the critical time period of treatment that has greatest effect on the outcome, or the long-term effect of some risk factors. Time-varying causal studies have recently emerged as a popular research area, leading to the development of numerous models and methods tailored to time-related questions. A recent review of time-varying causal models and methods can be found in the paper \cite{pazzagli2018methods}.

Though time-varying causal models and methods are numerous, they are hard to tailor to MR analysis. Many of the methods (like those based on structural nested mean models \cite{robins1989analysis,robins1994correcting} or marginal structural mean models \cite{robins2000marginal,michael2023instrumental}) rely on detailed individual-level data (e.g., comprehensive longitudinal measurements in one sample) and strong assumptions \cite{swanson2017nature}, and therefore, they may not be generally applicable to common real datasets. Most existing MR approaches for time-varying analysis consider the exposure at specific time points or overlook the time factor, possibly failing to provide a reliable understanding of the time-varying effects. Since the genetic variants are randomized at conception and their effects on the exposure are long-term rather than short-term fluctuations \cite{smith2006randomised}, MR estimates are usually interpreted as lifelong effects \cite[ch. 6.1.1]{burgess2021mendelian}. However, the lifelong effect usually represents a cumulative effect and cannot provide time-varying information during particular time periods. The challenges in time-varying MR studies have been discussed in a recent paper \cite{power2023methodological}.

In recent years, a growing body of literature has focused on the exploration of time-varying effects using MR. This includes work that considers repeated exposure measurements at different time points as separate exposures in a multivariable MR (MVMR) model \cite{richardson2020use,shi2021mendelian,richardson2022time,sanderson2022estimation}. However, these methods rely on a model represented by a directed acyclic graph (DAG) where each node corresponds to the phenotype at a specific timepoint—potentially predetermined by researchers. However, there is an absence of a clear and reasonable data-generating mechanism supporting the models in the time-varying context. Using DAGs with discrete time points for time-varying causal inference has been criticized \cite{aalen2016can,aalen2012causality}. Additionally, there is difficulty in interpreting the effects from such models. In particular, it is not clear what intervention corresponds to the estimated parameters. A recent paper has illustrated that employing the same risk factor measured at a limited number of measurement times in MVMR can potentially lead to misleading results \cite{tianestimation}.

Another line of research has explored the interpretation of MR estimates with a time-varying exposure. For instance, recent work suggests that MR with a single exposure measurement can yield misleading estimates if interpreted as the effect of varying the exposure in a mature population \cite{labrecque2019interpretation,labrecque2020commentary}. Furthermore, in scenarios with multiple measurement timepoints for the exposure, MR may fail to provide a solid basis for reasonable causal null testing over certain time regions \cite{swanson2018causal}. These challenges arise from the limited trajectory information considered for exposures, which violates the IV exclusion restriction assumption as the exposure continuously exists over time.

When we have limited exposure information available, such as measurements at specific time points, MR can still serve as a valid method for estimating specific cumulative time-varying quantities. For example, when the genetic effects remain constant, MR results can be interpreted as representing the cumulative effect over an individual's lifetime \cite{labrecque2019interpretation}. However, for more detailed time-varying conclusions, we rely on changes over time in the effects of the genetic variants on the exposure, which are referred to as time-varying genetic-exposure associations \cite[ch.~8]{burgess2021mendelian}. Some evidence suggests that genetic associations with certain exposures can indeed vary over time \cite{labrecque2021age}, which bolsters the feasibility of studying time-varying effects.

Given that time is continuous, and the corresponding effect of an exposure over time is also continuous, we should model the time-varying effect as a functional objective and incorporate the time-varying exposure trajectory into the model. Since the functional objective is infinite-dimensional, we utilize functional principal component analysis to achieve dimension reduction by extracting low-dimensional factors from exposure trajectories that can sufficiently explain time-varying information. Then, we integrate these low-dimensional factors into a multivariable MR model. These multivariable MR estimates are used to back out the functional time-varying exposure effect.

The aim of this work is to explore a method for time-varying MR analysis that can be applied in general settings. Some key features of our method are: we treat the time factor and the effect defined over time as a functional objective and use continuous-time modelling to study the time-varying effect; and we do not rely on strong parametric assumptions for the exposure generation mechanism. We allow the exposure to be measured at different time points for every individual, and each individual can have sparse exposure measurements. We also consider practical issues relating to time-varying MR studies under weak instruments.

The manuscript is structured as follows. In Section 2, we introduce general time-varying models and discuss the corresponding IV assumptions. We also delve into the issues related to identification under different settings and the challenges surrounding identification when employing existing MR methods.
Section 3 demonstrates the method of using functional principal component analysis to build a multivariable MR model and estimate time-varying effects as a functional objective. 
In Section 4, we present simulation examples with various time-varying instrument-exposure and exposure-outcome model scenarios.
In Section 5, we provide an illustrative example applying our methodology to real data to study the effect of blood pressure on urea levels.
We conclude in Section 6 with discussions.
The code implementing our time-varying MR analysis is available at: \url{https://github.com/HDTian/TVMR}.

\section{Models}

\begin{figure}[tbp]  
	\centering
	\includegraphics[width=0.75\textwidth]{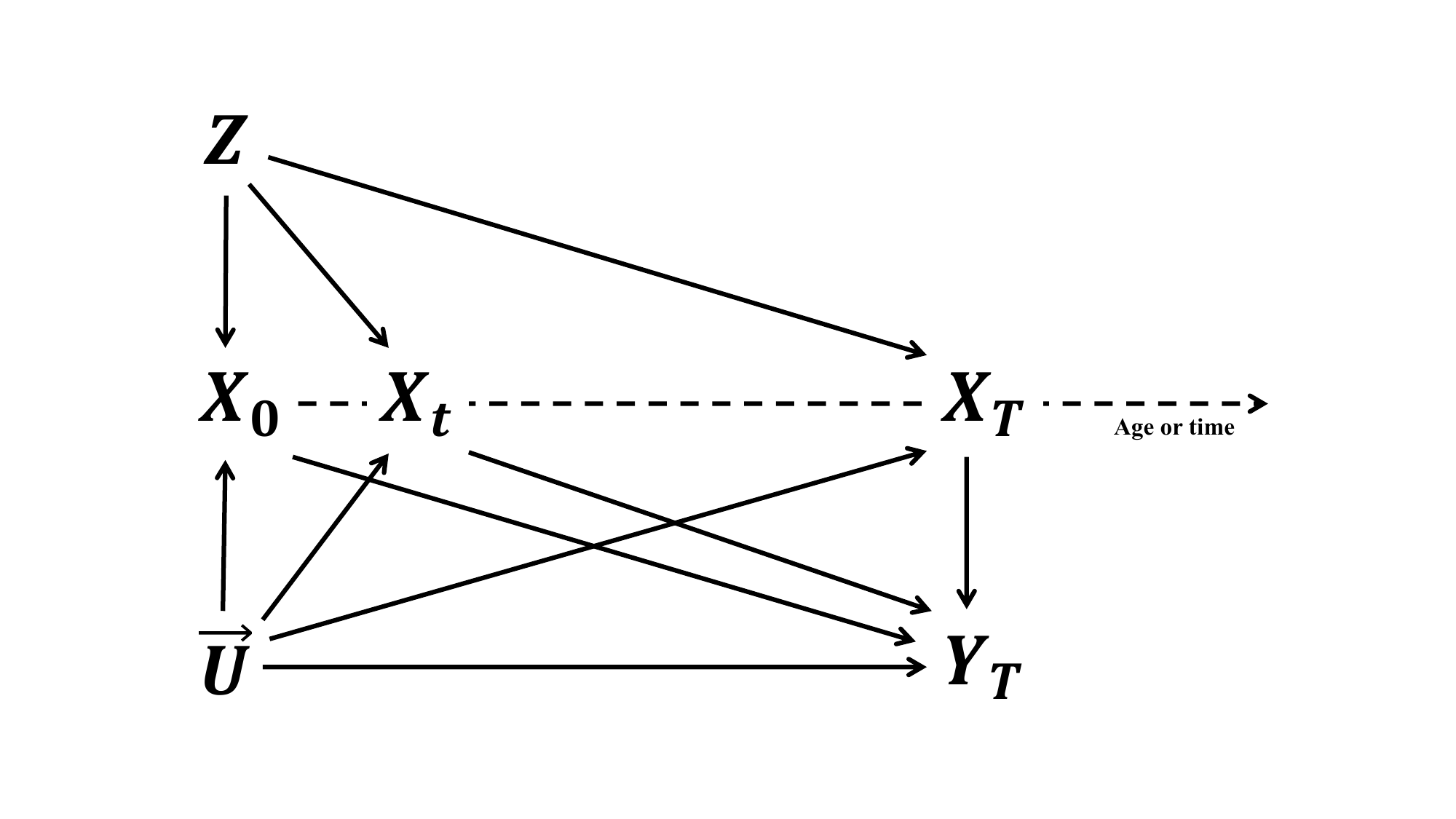} 
	\caption{The graph illustrates the IV assumptions for the time-varying Mendelian randomization setting. $X_0$, $X_t$, and $X_T$ represent the exposure at the time points $0$, $t$, and $T$, respectively, where $t$ is an arbitrary intermediate time point. $Z$ represents the time-invariant instrument existing at time point $0$ (e.g., the genetic variants). $\vec{U}=\{ U(t); 0 \leq t \leq T \}$ represents the process of the confounders. $Y_T$ is the outcome of interest, respectively. There should be no arrows from the exposure (or confounders) at a later time point to the outcome at present or past time points, as the future cannot influence the past.}
	\label{continuousDAG}
\end{figure}

We denote the time-varying exposure as a stochastic process starting from timepoint $s$ by $\vec{X}^s = \{X(t); t \geq s\}$, where $X(t)$ represents the exposure level at timepoint $t$. We let $\vec{X}=\vec{X}^0$. Furthermore, $Z$ and $U(t)$ represent a genetic variant and (possibly time-varying) unobserved confounders at timepoint $t$, respectively. 
The outcome of interest at timepoint $T$ is denoted as $Y_T$. The graphical representation of the relationships between these variables is illustrated in Figure \ref{continuousDAG}, which can be viewed as the continuous time version of the traditional instrumental variable DAG. We denote the potential outcome resulting from a complete exposure trajectory intervention $\vec{x}$ and an instrument intervention $z$ as $Y_T(\vec{x}, z)$. We assume that all potential outcomes are well-defined, adhering to the Stable Unit Treatment Value Assumption (SUTVA) \cite{rubin1980randomization}.

To investigate time-varying effects over the interval $[0, T]$ using instrumental variable (IV) methods, we require that any genetic variant used as an instrument satisfies the following three core IV assumptions:
\begin{itemize}
	\item[i.](Relevance) $cov(  Z ,  X(t) ) \neq 0$ for some $t$ over $[0,T]$  \label{IV1}  
	\item[ii.](Exchangeability) $ Z \indep Y_T( \vec{ x }, z )  $ for any $\vec{x}$ and $z$ \label{IV2}
	\item[iii.] (Exclusion restriction) $Y_T( \vec{ x }, z_1 )=Y_T( \vec{ x }, z_2 ) = Y_T(\vec{ x }) $ for any $z_1$, $ z_2$ and $\vec{ x }$\label{IV3}
\end{itemize}
Those assumptions are based on the potential outcome framework \cite{rubin1974estimating} and are analogous to the classical IV assumptions \cite{imbens2015causal}. However, in the time-varying case, the single exposure is replaced by the exposure trajectory, which is a functional variable depending on time. Relevance implies that each instrument should be associated with the exposure over some time regions. The exclusion restriction assumption states that each instrument can only affect the outcome through the exposure. Note that we do not require the exchangeability additionally satisfying $Z \indep \vec{X}(z)$, where $\vec{X}(z)$ represents the counterfactual exposure trajectory given the instrument level $z$. This means that the instrument can serve as a surrogate rather than a causal IV (e.g., a variant may be in correlation with some valid causal variants) \cite{hernan2006instruments}.

The three IV core assumptions are sufficient to test the joint sharp null hypothesis $H_0: \, Y_{T}(\vec{x}_1) - Y_{T}(\vec{x}_2) = 0$ for any individuals and $\vec{x}_1$ and $\vec{x}_2$, which means that the exposure has no effect on the outcome. However, such a conclusion cannot be further specified for specific time sub-regions. For example, to test the local sharp null hypothesis that $Y_{T}(\vec{x}^s_1) - Y_{T}(\vec{x}^s_2) = 0$ for an intermediate time point $0 < s < t < T$, the exclusion restriction originally satisfied for $[0, T]$ may be violated for $[s, T]$ (e.g., there may be a path $Z \to X_0 \to Y_T$ in Figure \ref{continuousDAG}). When the joint sharp null hypothesis is rejected, it is difficult to gain further insight on how a treatment effect may vary over time.

To investigate the time-varying effects beyond testing the joint sharp null, additional assumptions regarding the model are necessary. We assume the outcome is defined at the end timepoint $T$, and the structural equation for the outcome $Y$ is as follows:
\begin{equation}\label{model}
Y =  \beta_0+
\int_{0}^{T} \beta(t) X(t)dt+g_{Y}( \vec{U}, \epsilon_Y)
\end{equation}
Here, $g_{Y}(\cdot)$ is an unknown function, $\vec{U} = \{ U(t); 0 \leq t \leq T   \}$ is an unmeasured confounder trajectory, consisting of time-invariant variables and stochastic processes from the past up to the end time point. The term $\epsilon_Y$ denotes an exogenous error, and $\beta (t)$ represents the direct effect of the exposure at time point $t$ on the outcome, corresponding to the arrow from $X_t$ to $Y_T$ in Figure \ref{continuousDAG}. The function $\beta(t)$ provides complete information about the time-varying effect of the exposure on the outcome. Under any known time-varying treatment $\vec{a}$, one can express the effect of the exposure on the outcome via the counterfactual contrast
\begin{equation}
    \mathbb{E}( Y(  \vec{X} + \vec{a} )  -  Y(  \vec{X} )  ) = \int_0^T   \beta(t) a(t) dt
\end{equation}
We treat $\beta(t)$ as the functional objective of interest and call it the effect function.

We note that the discrete-time models used in recent studies \cite{richardson2020use,shi2021mendelian,richardson2022time,sanderson2022estimation} can be regarded as a special case of our time-varying model. For example, assuming $\beta(t) = \sum_{k=1}^{K} \beta_k \delta_D(t-t_k)$ for certain time points ${t_k, k=1,\ldots,K}$, where $\delta_D(t)$ represents the Dirac delta function.

We now take the inverse-variance-weighting (IVW) method as a simple estimation method, although other methods like the two-stage least squares (2SLS) lead to the same conclusion. Given the core IV assumptions, we can fit the following weighted regression: (see Supplementary Text S1)
\begin{equation}\label{fittingIVW}
	\hat{\theta}_j = \int_0^T \beta(t) \hat{\alpha}_j(t)dt + \epsilon_j \qquad \epsilon_j\sim \mathcal{N}(0, s.e.(\hat{\theta}_j)^2)\qquad j=1,2\ldots
\end{equation}
where $ \hat{\theta}_j $ represents the estimated association between the $j$-th instrument and the outcome, while $ \hat{\alpha}_j(t) $ represents the estimated association between the $j$-th instrument and the exposure at time point $t$. Equation \eqref{fittingIVW} captures the fundamental idea of time-varying MR studies: the identification of the time-varying effect $\beta(t)$ relies on the changes in exposure over time induced by the genetic variants (i.e., $\alpha_j(t)$) \cite[ch.~8]{burgess2021mendelian}. If $ \alpha_j(t) $ remains constant for any time point for all instruments, it becomes impossible to study the shape of $\beta(t)$. For instance, Labrecque \textit{et al.} demonstrated that in such cases, only the cumulative effect can be identified \cite{labrecque2019interpretation}. However, identifying the functional objective $\beta(t)$ necessitates some variation in shape among $\{\alpha_j(t), j=1,2,\ldots\}$ (i.e. the time-varying genetic-exposure association across the genetic instruments). We assume that the $\alpha_j(t)$ are sufficiently variable to allow estimation of the time-varying effect $\beta(t)$. A sufficient condition is that there is no timepoint when the $\alpha_j(t)$ are all zero, and there is no time period $t \in [t_1, t_2]$ where the $\alpha_j(t)$ are all constant or all linearly dependent.

An additional challenge arises from the high uncertainty typically associated with estimating the functions $ \hat{\alpha}_j(t)$. This issue is particularly prominent in time-varying studies where data are limited for certain time periods. To overcome this limitation and enhance power in time-related analyses, information may be borrowed across time.

\section{Methods}
Assume we can express any individual's exposure trajectory in a linear additive form as
\begin{equation}
    X(t) = \mu( t )  +  \sum_{k=1}^{K} \xi_{k}\phi_k( t )  + e(t)
\end{equation}
where the functional terms $ \{  \mu(t), \phi_k(t)  \} $ are known time-varying trends that are common for all individuals; the nonfunctional terms $ \{ \xi_k \} $ are the individual-specific variables, which summarize the variation in the exposure over time through $\phi_k(t) $; $e(t)$ is a time-varying error term for the exposure and assumed to be sufficiently small so that the structural model \eqref{model} becomes
\begin{equation}\label{MVMR}
\begin{split}
Y_i =& \beta_0+ \int_{0}^{T} \beta(t) X_i(t)dt + g_Y(U_i,\epsilon_{Y,i}) \\
=&   \beta_0+ \int_{0}^{T} \beta(t)\left[\mu( t )  +  \sum_{k=1}^{K} \xi_{k,i}\phi_k( t )   + e_i(t) \right] dt + g_Y(U_i,\epsilon_{Y,i})  \\
=& \underbrace{ \left[  \beta_0+ \int_{0}^{T} \beta(t)\mu(t)dt    \right] }_{\beta_0^\ast}+ \sum_{k=1}^K\underbrace{ \left[\int_{0}^{T} \beta(t)\phi_k( t )dt \right]}_{   \beta_k^\ast} \xi_{k,i}+\underbrace{  \int_0^T \beta(t) e_i(t)dt   + g_Y(U_i,\epsilon_{Y,i})}_{\epsilon_i}
\end{split}
\end{equation}
which is the classical form of MVMR where $\{\xi_k\}$ are considered the exposure variables, and their effects are given by $\{\beta^\star_k\}$. We refer to $\{\xi_k\}$ as pseudo-exposures.

Using the linear additive form for the exposure trajectory reduces the original infinite-dimensional problem with respect to the effect function $\beta(t)$ to a $K$-dimensional problem with respect to the components $\{\xi_k \} $ that can be addressed through IV analysis. The remainder term $ \int_0^T \beta(t) e_i(t)dt $ may lead to a violation of the IV core assumptions, especially when the trajectory approximation is poor. We refer to this as structural pleiotropy, which can be negligible if $e(t) $ is sufficiently small or is independent of the instruments.

Several methods exist to accomplish such dimension reduction. One approach involves making a parametric assumption for individual exposure trajectories and extrapolating or interpolating the exposure for each individual. Alternatively, one can first estimate the genetic association with the exposure $\hat{\alpha}(t) $ at multiple timepoints and then smooth these values over time. However, these methods come with their own set of challenges. The former method relies on parametric assumptions. The latter method is likely to suffer from high estimate uncertainty for any specific timepoint due to the limited sample size for the exposure at certain measured timepoints. This motivates us to consider more efficient dimension reduction methods to avoid these problems.

\subsection{Estimating $\beta(t)$}\label{Estimating}
Consider the functional approximation \eqref{MVMR}. In the case that $ cov( \xi_k, \epsilon ) =0$ for any $k$, we can estimate the effect shape $ \beta(t)$ and its confidence intervals by the following two steps:
\begin{itemize}
	\item[1.] First, fit linear regression of $Y$ on the pseudo-exposures $\{ \xi_k \}$ to obtain the estimates $ \hat{ \boldsymbol{\beta} }^\ast $ and the covariance matrix $ \Sigma_{\boldsymbol{\beta}^\ast} $.
 \item[2.] Invert the equation $ \beta_k^\ast = \int \beta(t) \phi_k(t)dt $ using the functions $\phi(t)$ and plug in the estimates $ \hat{\beta}^\ast_k $ (as well as $ \Sigma_{\hat{\boldsymbol{\beta}}^\ast} $) to obtain $ \hat{ \beta }(t) $. Section \ref{Curve} describes further details of inverting the integral equation $ \beta_k^\ast = \int \beta(t) \phi_k(t)dt $.
\end{itemize}
Since the exposure trajectory $ \{ X(t)\}$ (and thus $ \{ \xi_k \}$) can be confounded with the outcome $Y$, $ cov(\xi_k, U) \neq 0  $ and $ cov(   \xi_k , \epsilon ) = 0 $ may not hold, and therefore we instrument the pseudo-exposures $\xi_k$ using genetic variants.

\subsection{Multi-principal-components Mendelian randomization}
In this section, we will introduce (i) how to apply data-driven methods for functional dimension reduction, (ii) using instruments to study time-varying effects, and finally, (iii) how to derive the effect function estimator.

\subsubsection{Functional principal component analysis}
One effective approach to achieve dimension reduction for the exposure trajectories and reduce the variance of the remainder term $e(t)$ is called functional principal component analysis (FPCA). The Karhunen--Loève theorem states that, under some regularity conditions,
$
		X(t) = \mu( t )  +  \sum_{k=1}^{\infty} \xi_{k}\phi_k( t )
$
where $\mu(t)=\mathbb{E}( X(t) )$, $ \xi_{k} $ represents the principal component (PC) value with respect to the $k$-th eigenfunction $\phi_k(t)$, and $ \lambda_k:=\text{var}( \xi_k ) $ denotes the eigenvalues such that $ \lambda_1>\lambda_2>\cdots $. FPCA transforms the variable trajectories into a PC-weighted sum of eigenfunctions, ordered by the primary PC scores that explain more variance. For more details, please refer to Supplementary Text S2. Therefore, we can approximate the exposure trajectory using the first $K$ PCs that explain a sufficient amount of variation, making $e_i(t)$ sufficiently small. In practice, the choice of $K$ can be based on the number of principal components required to explain a certain proportion of exposure variation. Since the available data may have only sparse measurement time points and the exposure measurements may be observed at different time points for different individuals, one method for conducting FPCA in such scenarios is principal component analysis through conditional expectation (PACE) \cite{yao2005functional,wang2016functional}. PACE helps to estimate individual PC scores $ \xi_{i,k} $ and the eigenfunctions $\phi_k(t) $ based on the properties of FPCA. Further details can be found in Supplementary Text S3.

\begin{figure}[tbp]  
	\centering
	\includegraphics[width=0.8\textwidth]{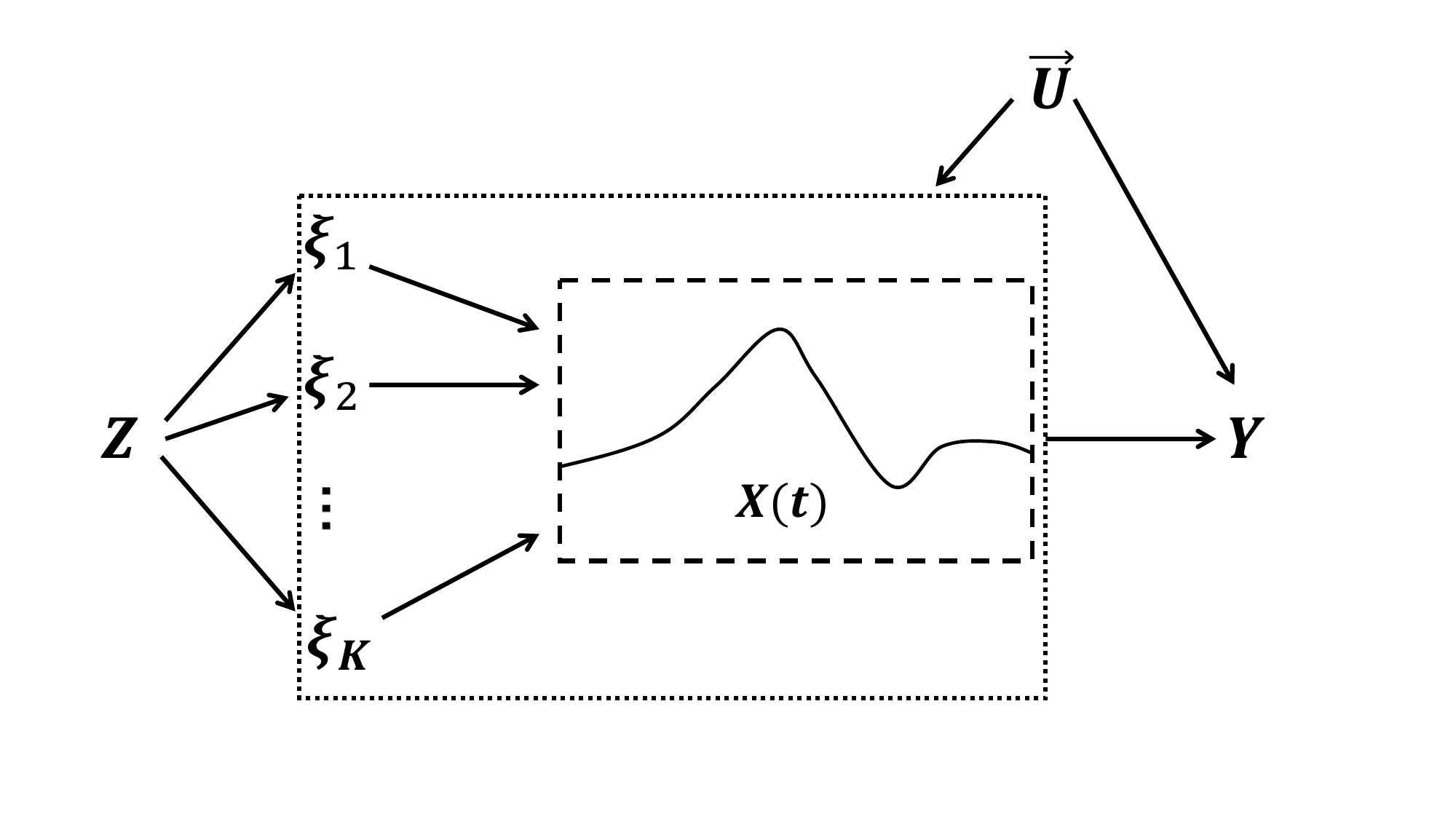} 
	\caption{The graph of a Mendelian randomization analysis under the MPCMR model. $Z$ represents the high-dimensional genetic instruments, $ {\xi_1,\ldots,\xi_K} $ are the $K$ principal components associated with the individual exposure curve $ {X(t)} $, $\vec{U}=\{U(t);0 \leq t\leq T\}$ stands for unmeasured confounders, and $Y$ denotes the outcome. }
	\label{DAG}
\end{figure}

The analogous diagram illustrating Multi-principal-components Mendelian randomization (MPCMR) is provided in Figure \ref{DAG}. In this graph, the pseudo-exposures $ \{ \xi_1,\ldots,\xi_K \}$ can be interpreted as latent time-invariant variables mediating the path from the genetic variants to the time-varying exposure   \cite{morris2022interpretation,sanderson2022estimation}. 

\subsubsection{MPCMR fitting}
Once we have computed the principal components (PCs), we can fit the MPCMR models as described in \eqref{MVMR} using standard IV methods. We can estimate the effect function using the same two-step procedure as in the association model described in Section \ref{Estimating}. We consider the continuously updating generalized method of moments (GMM) \cite{hansen1996finite} rather than the IVW method for estimation, as it is less sensitive to weak instruments \cite{antoine2009efficient}. To proceed, assume that all variables (including the genetic variants, principal components, and the outcome) have been mean centered. We denote the sample mean of any random variable $X$ with a sample size of $n$ as $ \hat{ \mathbb{E} }_n( X ) $. We define the $J$-vector of instruments as $\boldsymbol{Z}  $ and the $K$-vector of pseudo-exposures (principal components) as $ \boldsymbol{\xi} $. The GMM estimator is:
\begin{equation}
    \hat{\boldsymbol{\beta}}^\ast= \argmin_{ \boldsymbol \beta } \hat{\boldsymbol g}( \boldsymbol \beta) \hat{\Omega }(\boldsymbol \beta)^{-1} \hat{\boldsymbol g}( \boldsymbol \beta )
    \end{equation}
where $ \hat{\boldsymbol g}( {\boldsymbol \beta} ) =   \hat{ \mathbb{E}}_n( \boldsymbol{Z} ( Y-\boldsymbol{\xi}^T {\boldsymbol{\beta}} )  )  $ and $ \hat{ \Omega }(  \boldsymbol{\beta}  ) = \hat{\mathbb{E}}_n(   \boldsymbol{Z} \boldsymbol{Z}^T )  \hat{\mathbb{E}}_n(  ( Y- \boldsymbol{\xi}_i^T {\boldsymbol{\beta}} )^2 )  $. Under the assumption that the covariance between the composite error term $\epsilon$ in Equation \eqref{MVMR} and the instruments $\boldsymbol{Z}$ is assumed to be zero, homoskedastic errors, and standard GMM regularity conditions \cite{newey1994large}, we have
\begin{equation}
    \sqrt{n} (  \hat{ \boldsymbol\beta }^\ast  -  \boldsymbol{\beta}^\ast ) \overset{D}{\to} \mathcal{N}(  \boldsymbol{0} , (\boldsymbol{G}^T \Omega( \boldsymbol{\beta}^\ast )^{-1}  \boldsymbol{G}    )^{-1})
\end{equation}
where $  \boldsymbol{G} = \mathbb{E}(  - \boldsymbol{Z} \boldsymbol{\xi}^T ) $ and $ \boldsymbol{\Omega}( \boldsymbol{\beta}^\ast ) = \plim_{n \rightarrow \infty} \hat{ \Omega }(  \boldsymbol{\beta}^\ast  ) $. We can obtain the estimated variance-covariance matrix of $ \hat{\boldsymbol{\beta}}^\ast $, denoted by $  \hat{ \Sigma } $, with plug-in estimates $ \hat{ \boldsymbol\beta }^\ast   $ and $ \hat{ \boldsymbol{G} } = \hat{\mathbb{E}}_n(  - \boldsymbol{Z} \boldsymbol{\xi}^T ) $. That is, $ \hat{ \Sigma } =   ( \hat{\boldsymbol{G}}^T \hat{\Omega}( \hat{\boldsymbol{\beta}}^\ast )^{-1}  \hat{\boldsymbol{G}}    )^{-1} $.

\subsubsection{Curve inversion}\label{Curve}
We can proceed to invert the equation $ \beta_k^\ast = \int \beta(t) \phi_k(t)dt $ to obtain the functional objective $ \beta(t) $. Define $  \boldsymbol{\phi} = ( \phi_1(t) \, \phi_2(t) \, \ldots \, \phi_K(t) )^T $. Given that $ \boldsymbol{\beta}^\ast = \int \boldsymbol{\phi}(t) \beta(t) dt $ and the eigenfunction property $ \int \boldsymbol{\phi}(t) \boldsymbol{\phi}^T(t) dt = \boldsymbol{I} $, we can immediately find one solution for the equation, which is $ \beta(t) = \boldsymbol{\phi}^T(t) \bold{ \boldsymbol{\beta}^\ast} $. Therefore, we have one final MR estimator for the time-varying effect function:
\begin{equation}\label{eq7}
    \hat{ \beta }(t) = \boldsymbol{\phi}^T(t)\hat{  \boldsymbol{\beta} }^\ast = \sum_{k=1}^K  \phi_k(t)\hat{\beta}^\ast_k
\end{equation}
and the pointwise standard errors $  \hat{\sigma}(t)= \sqrt{ \boldsymbol{\phi}^T(t)  \hat{\Sigma}  \boldsymbol{\phi}(t) }  $.

Note that the inversion of $ \boldsymbol{\beta}^\ast $ presents an ill-posed inverse problem. This means that the time-varying effect $\beta(t)$ may not be identified even if we have a consistent estimator of $ {\boldsymbol{\beta}}^\ast$ \cite{horowitz2014ill}. There exist infinitely many solutions for $ \beta(t) $ such that the equation $ \boldsymbol{\beta}^\ast = \int \boldsymbol{\phi}(t) \beta(t) dt $ hold.
To address this issue, additional parametric assumptions may be useful to study the objective function \cite{newey2003}. These assumptions can help constrain the possible solutions and improve the identifiability of the time-varying effect.

In the general case, we assume that the time-varying function can be represented as a linear additive combination of basis functions, such that $ \beta(t) = \sum_{l=1}^L \gamma_l b_l(t) $. Here, $ \{ b_l(t)\} $ are known basis functions, and $ \{\gamma_l \}$ are the parameters of interest representing the contribution of each basis function to the time-varying effect. The structural outcome equation can be rewritten to accommodate this parametric form as follows:
\begin{equation}
\begin{split}
    Y_i =&   \beta_0+ \sum_{k=1}^K \left[\int_{0}^{T} \beta(t)\phi_k( t )dt \right] \xi_{k,i}+ \epsilon_i  \\
     = & \beta_0+ \sum_{k=1}^K \left[\int_{0}^{T} \left[  \sum_{l=1}^L \gamma_l b_l(t)  \right]   \phi_k( t )dt \right] \xi_{k,i}+ \epsilon_i  \\
     =& \beta_0+  \sum_{l=1}^L  \gamma_l \underbrace{ \left[  \sum_{k=1}^K \int_{0}^{T}  b_l(t) \phi_k( t )dt \, \xi_{k,i} \right] }_{ \xi^\ast_{l,i} }+ \epsilon_i
    \end{split}
\end{equation}
where we refer to $ \boldsymbol{\xi}^\ast \in \mathbb{R}^{L\times 1} $ as the transformed exposure (principal components) vector, which satisfies $ \boldsymbol{\xi}^\ast = \boldsymbol{B}^T\boldsymbol{\xi} $, where $ \boldsymbol{ B } = \int_{0}^{T} \boldsymbol{ \phi }(t) \boldsymbol{ b }^T(t) dt \in \mathbb{R}^{K\times L} $ is a known transforming matrix. Therefore, we can fit the MPCMR with the exact same procedures, but replace the exposures with the transformed exposures. In GMM, we can simply replace the $K$-vector of pseudo-exposure with the $L$-vector of transformed exposure. Note that $ L \leq K $ is required to achieve the identification of the parameters $ \{\gamma_l \}$. 

\begin{figure}[tbp]  
	\centering
	\includegraphics[width=0.8\textwidth]{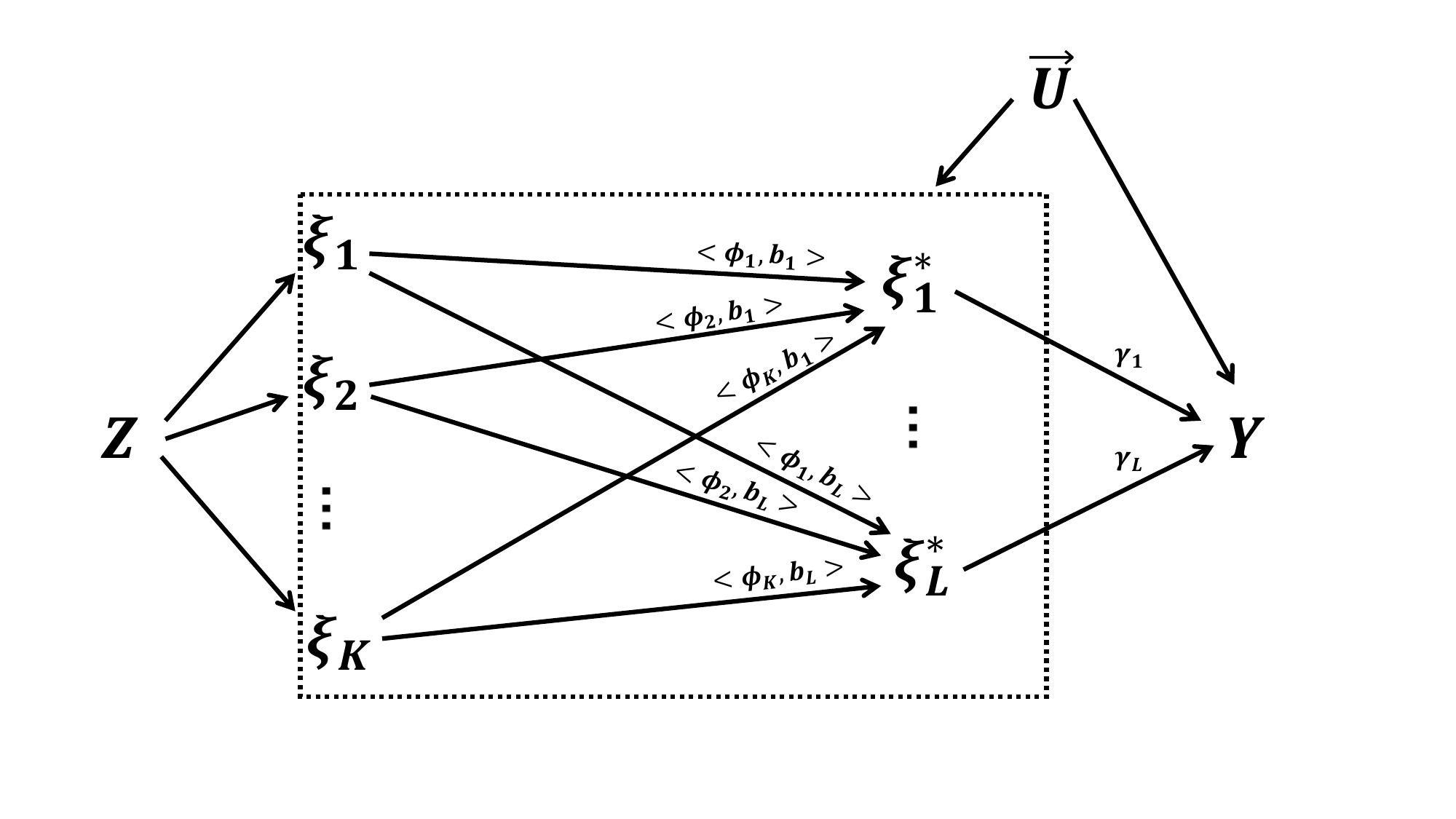} 
	\caption{The diagram of Mendelian randomization under the MPCMR model with transformed principal components. $Z$ represents (high-dimensional) genetic instruments, ${\xi_1,\ldots,\xi_K}$ are the original $K$ principal components for the individual exposure curve, ${\xi^\ast_1,\ldots,\xi^\ast_L}$ are the transformed $L$ principal components, $\vec{U}=\{U(t); 0 \leq t \leq T\}$ represents unmeasured confounders, and $Y$ is the outcome. $< \phi_k,b_l > = \int \phi_k(t)b_l(t) dt$ for any $k$ and $l$, where $\phi_k(t)$ is the $k$-th eigenfunction and $b_l(t)$ is the $l$-th known basis function. ${\gamma_1 , \ldots, \gamma_L }$ are the parameters of interest.}
	\label{DAGt}
\end{figure}

The final effect function is calculated as follows:
\begin{equation}
    \hat{\beta}(t) =\boldsymbol{b}^T(t)\hat{  \boldsymbol{\gamma} } =  \sum_{l=1}^L b_l(t) \hat{ \gamma }_l
\end{equation}
where the estimates $ \hat{ \boldsymbol{\gamma} } $ represent the MPCMR estimates with transformed principal components. The pointwise standard error is denoted as $ \hat{\sigma}(t) = \sqrt{ \boldsymbol{b}^T(t) \hat{\Sigma} \boldsymbol{b}(t) } $. The MVMR model with the general transformed principal components is demonstrated in Figure \ref{DAGt}.

Note that one special example is to let $ \boldsymbol{b}(t) = \boldsymbol{\phi}(t) $; that is, the eigenfunction is also the basis function. According to the eigenfunction property, $ \boldsymbol{B} = \boldsymbol{I} $, therefore the transformed principal components are the principal components themselves. In addition to the eigenfunctions, another classical choice of the basis functions are the polynomial basis functions $ b_l(t) = t^{l-1} $ for $l=1,2,\ldots$. 

The approach we consider could be implemented in a two-sample setting, where genetic variant associations with the exposure are measured in a non-overlapping, but representative, sample to genetic variant associations with the outcome, and with summary-level data, given that summary statistics relating to the principal components are derived. The details for these settings are provided in Supplementary Text S4-S5.

\subsection{Time-varying IV strength analysis}
The usual methods for assessing instrument relevance and validity can be used for our time-varying MR models. For example, we can assess the evidence on whether all the instruments used are valid using heterogeneity or over-identification tests \cite{hansen1982large}. Similarly, we can evaluate instrument strength to determine if the instrument can support meaningful and reliable time-varying results. The IV strength or validity assessment for time-varying MR is similar to those for MVMR, where the principal components are treated as the exposures. Details are given in Supplementary Text S6-S7.

Instrument strength for the pseudo-exposure $\xi$ is closely related to instrument strength for the time-varying exposure $X(t)$. Instead of measuring the strength of genetic associations with the exposure at multiple time points, the time-varying association can be evaluated in a more efficient way using the FPCA approximation:
\begin{equation}\label{GX}
    \alpha(t)= \frac{cov( G, X(t) )}{ var( G ) } \approx  \frac{cov( G, \sum_{k=1}^K \xi_k \phi_k(t) )}{ var( G ) } = \sum_{k=1}^{K} \frac{cov(G,\xi_k)}{ var(G) } \phi_k(t) = \sum_{k=1}^{K} {\alpha_k}\phi_k(t)
\end{equation}
where $\alpha_k$ represents the genetic association with the $k$-th principal component. Therefore, genetic associations with the time-varying exposure can be expressed as a weighted average of genetic associations with the principal components. Instrument strength in multivariable MR is measured by conditional F-statistics \cite{sanderson2016weak} (or Cochran's Q statistic \cite{sanderson2019examination}, both embedded in our package TVMR).

\subsection{Identification-robust MR analysis}
If the time-varying genetic associations with the exposure are be similar (i.e., approximately proportional) to each other, the genetic associations with the principal components would be approximately proportional. This would lead to the genetic variants being conditionally weak instruments. If we increase the number of principal components used to model the exposure, instrument strength is likely to get weaker. Therefore, we consider identification-robust inference, to ensure that the probability of falsely rejecting the null hypothesis can be well-controlled regardless of whether the instrument is weak or the model is ill-identified \cite{andrews2019weak}.
Identification-robust methods do not assume the exposure effects are point-identifiable and hence do not return point estimates. Importantly, they return valid confidence intervals even under weak instrument settings. In contrast, inferential methods based on IVW or GMM point estimates can be biased under weakly identified exposure effects.
There are many statistics proposed for identification-robust inference (see \cite{andrews2019weak}). As a popular statistic that can also work for over-identified models, Kleibergen's Lagrange multiplier (LM) statistic \cite{kleibergen2005testing} is used in this paper. We provide a derivation to calculate and perform inferences on this statistic based on summarized genetic association data.

Following usual GMM arguments \cite{kleibergen2005testing}, we have the joint asymptotic distribution
\begin{equation}
\sqrt{n}
\left( \begin{array}{c}
\hat{ \boldsymbol{g} }( \boldsymbol{\beta}^\ast )  \\
\textrm{vec}(  \hat{ \boldsymbol{G} } )
\end{array} \right)  \overset{D}{\to}   \mathcal{N}\left(
\left( \begin{array}{c}
\boldsymbol{0} \\
\boldsymbol{\mu}
\end{array} \right)  ,
\left( \begin{array}{cc}
\Omega( \boldsymbol{\beta}^\ast ) & \Delta \\
\Delta^T & \Sigma_{    G }
\end{array} \right)		\right)
\end{equation}
where $ \text{vec}( \hat{ \boldsymbol{G} } )$ is the $JL$-vector obtained by stacking the columns of $ \hat{ \boldsymbol{G} } $. 
We let $\Delta \in \mathbb{R}^{J \times JL}$ denote the covariance matrix of $ \hat{ \boldsymbol{g} }( \boldsymbol{\beta}^\ast )   $, and let $\Sigma_G \in \mathbb{R}^{JL \times JL} $ denote the the covariance matrix of $ \textrm{vec}(  \hat{ \boldsymbol{G} } )$.
The vector $\sqrt{n}\textrm{vec}(\hat{\boldsymbol{G}}  )$ has a fixed mean vector $\boldsymbol \mu$ due to the weak instrument assumption that the covariance between any exposure (principal component) and instruments is decreasing at the same rate as the sampling error $1/\sqrt{n} $.
  Under the null hypothesis $ H_0: \boldsymbol{\beta}^\ast = \boldsymbol{\beta}_0 $, the weak-instrument-robust approach aims to construct a test statistic (LM statistic) based on a sufficient statistic for $\hat{\boldsymbol{G}}$ that is asymptotically uncorrelated with $\hat{\boldsymbol{g}}( \boldsymbol{\beta}_0 ) $ under $H_0$. The calculation of LM statistic under $H_0$ involves three steps (for more details, see the supplementary text S8):
\begin{itemize}
	\item[1.] We construct a consistent estimator of $\Delta$. The $k$-th column of $\hat{\boldsymbol{G}}$ is $\hat{\boldsymbol G}_k=  \hat{\mathbb{E}}_n( - \boldsymbol{Z} \xi_k )   $. Denote $ \Delta= (  \Delta_1 \, \Delta_2 \, \ldots \, \Delta_K ) $ where $ \Delta_k := cov(
 \sqrt{n} \hat{ \boldsymbol{g} }( \boldsymbol{\beta}_0 ) , \sqrt{n} \hat{\boldsymbol G}_k )  $. We can construct consistent estimators $ \hat{\Delta}_k=\hat{ \mathbb{E} }_n(  - \boldsymbol{Z} \xi_k  (Y - \boldsymbol{\xi}^T \boldsymbol{\beta}_0) \boldsymbol{Z}^T  ) $ and $ \hat{ \Delta} = (  \hat{\Delta}_1 \, \ldots \, \hat{\Delta}_K ) $.

    \item[2.] Then, for each principal component k, let $  \hat{D}_k( \boldsymbol{\beta}_0 ) = \hat{\boldsymbol{G}}_k - \hat{\Delta}_k(  \boldsymbol{\beta}_0 )^T \hat{\Omega}(\boldsymbol{\beta}_0)^{-1} \hat{\boldsymbol{g}}( \boldsymbol{\beta}_0 ) $. A sufficient statistic of $\hat{\boldsymbol{G}} $ is $ \hat{\boldsymbol D}( \boldsymbol{\beta}_0 ) = (  \hat{D}_1 (\boldsymbol{\beta}_0) \,  \ldots \,  \hat{D}_K (\boldsymbol{\beta}_0)   )$; it can be shown that each principal component $k$, $\hat{D}_k( \boldsymbol{\beta}_0 )$ is asymptotically uncorrelated with $\hat{\boldsymbol{g} }( \boldsymbol{\beta}_0 ) $.

    \item[3.] Finally, the LM statistic is
\begin{equation}
    LM( \boldsymbol{\beta}_0 ) = [      \hat{ \Omega}( \boldsymbol{\beta})_0^{-\frac{1}{2}}  \sqrt{n}  \hat{ \boldsymbol{g} }(  \boldsymbol{\beta}_0   )          ]^T  \hat{\boldsymbol{P}}( \boldsymbol{\beta}_0 ) [           \hat{ \Omega}( \boldsymbol{\beta}_0)^{-\frac{1}{2}}  \sqrt{n}  \hat{ \boldsymbol{g} }(  \boldsymbol{\beta}_0   )                  ]
\end{equation}
where $  \hat{\boldsymbol{P}}( \boldsymbol{\beta}_0 ) = [ \hat{ \Omega}( \boldsymbol{\beta})_0^{-\frac{1}{2}} \hat{\boldsymbol{D}}( \boldsymbol{\beta}_0 )       ]
[           \hat{\boldsymbol{D}}( \boldsymbol{\beta}_0 )^T   \hat{\Omega}( \boldsymbol{\beta}_0 )^{-1}    \hat{\boldsymbol{D}}( \boldsymbol{\beta}_0 )              ]^{-1}
[ \hat{ \Omega}( \boldsymbol{\beta})_0^{-\frac{1}{2}} \hat{\boldsymbol{D}}( \boldsymbol{\beta}_0 )       ]^{T}$
\end{itemize}
Under $H_0: \boldsymbol{\beta}^\ast = \boldsymbol{\beta}_0$ and some regularity conditions, Theorem 1 of  \cite{kleibergen2005testing} shows that $ LM(\boldsymbol{\beta}_0)  \overset{D}{\to} \chi^2_K $. Thus, confidence regions for $\beta^\star$ can be obtained by inverting the LM test. In particular, one can follow these steps:
(i) Assign $m$ increasing points for each parameter, resulting in $m^K$ candidate values for $\boldsymbol{\beta}_0$;
(ii) Calculate the LM statistic over each candidate value;
(iii) Collect the parameter values for $\boldsymbol{\beta}^\star$ that are not rejected by an LM test.
Candidates within the confidence region for $\boldsymbol{\beta}^\star $ can be further used to obtain the confidence interval for the effect function $ \beta(t)$.

\section{Simulation}
\subsection{Simulation design}
The aim of our simulation study is to investigate the ability of the MPCMR approach to estimate different forms of time-varying exposure effects.
We consider different scenarios for the structural exposure and outcome models. Let the time region of interest be $[0,50]$. For the exposure, we consider the following model:
\begin{equation}   \label{exposure_model}
X(t) =\underbrace{ \sum_{j=1}^{30} \alpha_j(t) G_j}_{\textrm{gene score}}  + \underbrace{[U_0+ U(t)]}_{\textrm{confounding effect}} + \epsilon_X(t)    \qquad t\in[0,50]  
\end{equation}
where $ G_j $ represents the $j$-th genetic variant and $\alpha_j(t)$ is the time-varying genetic effect on the exposure.
Genetic variants were generated as $G_j \overset{\textrm{i.i.d}}{\sim}B(2,0.3)$. The terms $ U(t) $ and $\epsilon_X(t)  $ were set to be independent Wiener processes with variance equal to $1$ at $t=50$, and $ U_0 \sim \mathcal{N}(0,1^2) $ is a time-fixed confounder. We consider   three functions for the instrument--exposure relationships, a sinusoidal function (A) and two linear functions (B and C): 
\begin{enumerate}
	\item[A.]  $  \alpha_j(t)= 0.05 \sin{  (a_j t)   } + b_j $ where $ a_j \overset{\textrm{i.i.d}}{\sim} Unif(-0.1,0.1) $ and $ b_j \overset{\textrm{i.i.d}}{\sim} Unif(-0.1,0.1) $
	
	\item[B.] $\alpha_j(t)= a_j + b_j t$ where $ a_j \overset{\textrm{i.i.d}}{\sim} Unif(-0.1,0.1) $ and $ b_j \overset{\textrm{i.i.d}}{\sim} Unif(-0.004,0.004) $

    \item[C.] $\alpha_j(t)= a_j + b_j t$ where $ a_j \overset{\textrm{i.i.d}}{\sim} Unif(-0.1,0.1) $ and $ b_j \overset{\textrm{i.i.d}}{\sim} Unif(-0.01,0.01) $
\end{enumerate}

\begin{figure}[tbp]
	\centering
	\includegraphics[width=0.7\textwidth]{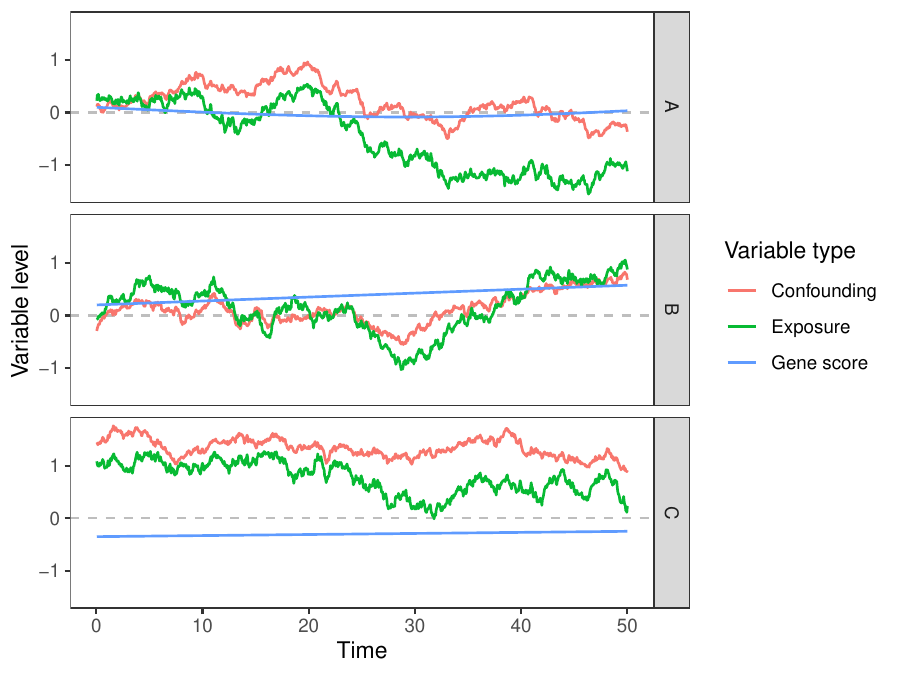} 
	\caption{Examples of the individual variables for the exposure, gene score, and confounding curves over time under three different instrument-exposure models A-C.}
	\label{mf00}
\end{figure}
The instrument strength for the exposure at any single timepoint is controlled for scenarios A-B such that the average proportion of variation explained by genetic variants is approximately $0.05$. We also control the genetic time-varying effects so that different scenarios have different average instrument strength measurements in the time-varying MR context. This will help to investigate the performance of our method under different IV strengths.
Figure \ref{mf00} shows examples of the time-varying variables under different scenarios. We selected $K=2$ principal components that explained at least 95\% of exposure variation. The corresponding first two eigenfunctions are of an approximately linear form. We consider two sets of basis functions for estimating the time-varying effect of the exposure on the outcome: the eigenfunctions, and polynomial functions of degree 1 (i.e. linear functions). Fitting more complex exposure--outcome models would require more principal components to be estimated. However, this would lead to weaker instruments. We return to discuss the trade-off between model complexity and instrument strength in Section \ref{Discussion}.

We consider the outcome model
\begin{equation}
Y=   \int_0^{50} \beta(t) X(t) dt  +10[U_0 +  U(50)] + \epsilon_Y(50)
\end{equation}
We consider the following six exposure-outcome models with different causal shapes $\beta(t)$, denoted by Scenario 1-6:
\begin{enumerate}
	\item[1.] Lifetime null effect: $\beta(t)\equiv 0$
	
	\item[2.] Lifetime constant effect: $\beta(t)\equiv 0.1$
	
	\item[3.] Lifetime increasing effect: $ \beta(t)=0.02 t $
	
	\item[4.] Lifetime sign-change effect:  $ \beta(t)=0.5- 0.02 t $

    \item[5.] Early-age-only effect: $ \beta(t)=0.05(-t+20) I\{t<20\} $
	
    \item[6.] Later-age-only effect: $ \beta(t)=0.05( t-30 ) I\{t>30\} $

\end{enumerate}
Scenario 1 is the joint sharp causal null case \cite{swanson2018causal}. Scenario 2 is the constant effect case, which means the intervention of the exposure at any timepoint will have the same effect on the outcome. Scenario 3 indicates that the exposure has an increasingly stronger effect on the outcome over age, which means the later the intervention of exposure is given, the greater the effect on the outcome. Scenario 4 indicates that early-age exposure has a positive effect but later-age exposure has a negative effect. Scenarios 5 and 6 are threshold effect cases where there exists a certain period under which intervening on the exposure influences the outcome.
Note that the outcome models 1-4 are of a linear form, which can be well captured by the polynomial basis function as well as the eigenfunctions (since the first two eigenfunctions are approximately linear). However, the non-smooth effect functions in scenarios 5-6 cannot be well approximated by the polynomials or eigenfunctions. Therefore, our approach is unlikely to perform well under Scenarios 5 and 6. Instead, one should consider other appropriate basis functions according to domain knowledge regarding the underlying shape of the effect function, or consider data-driven approaches to select relevant basis functions.

We consider $n=10,000$ individuals in the simulation and randomly select 10 time points for each individual from the region $[0, 50]$ to mimic a sparse measurement scenario, which is common in practice. We first calculate functional principal components using the widely-used package `fdapace' \cite{fdapace}. We then fit the time-varying model using three different strategies:
\begin{enumerate}
	\item[$\bullet$] Association: naive association fitting for the principal components and the outcome without using instruments;

    \item[$\bullet$] MPCMR(eigenfunction): MPCMR where the basis functions are the eigenfunctions;

    \item[$\bullet$] MPCMR(polynomial): MPCMR where the basis functions are linear functions.
\end{enumerate}
 We use the weak instrument-robust LM statistic to calculate confidence intervals for the fitted effect function for strategies (ii) and (iii). When building the confidence intervals with the LM statistic, for each parameter $ \beta^\ast_k$, we search over the parameter regions $[ \hat{\beta}^\ast_k - 4s.e.( \hat{\beta}^\ast_k ), \hat{\beta}^\ast_k + 4s.e.( \hat{\beta}^\ast_k ) ] $, which is sufficiently large to cover non-rejected regions in our simulation setting.

We simulate and fit the data $1,000$ times independently for each scenario and calculate the Mean Squared Error (MSE) of the GMM point estimates and coverage for the effect function rate at $4$ time points ($t=10,20,30,40$). As is common in functional data analysis, we avoid comparisons over time points near the boundaries (e.g., $t=0,50$), where the curve estimates tend to have much greater errors \cite[Page~68]{Ramsayfunctionaldata}.
To measure instrument strength, we calculate the conditional F-statistic relating to each principal component.

\subsection{Simulation results}
\begin{figure}[tbp]
	\centering
	\includegraphics[width=0.8\textwidth]{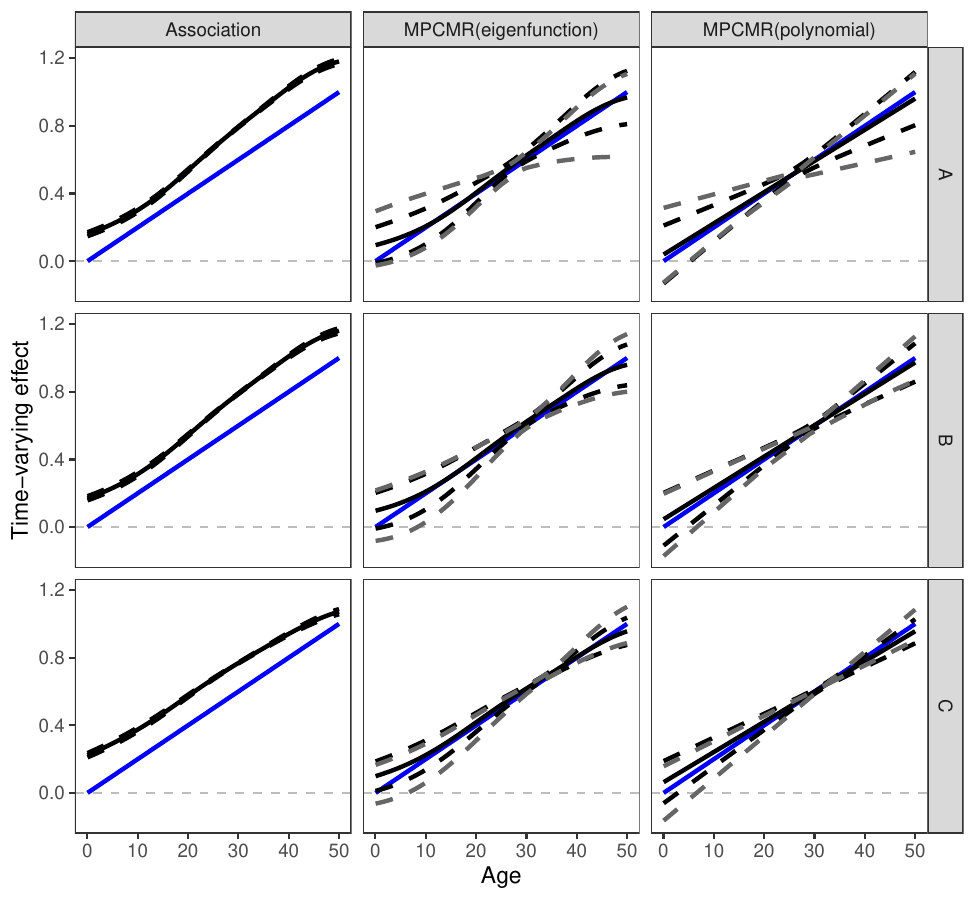} 
	\caption{Time-varying MR fitting results using different estimation strategies under different instrument-exposure models under one single simulated example. The rows correspond to the instrument-exposure models A, B, and C, respectively. The columns correspond to the estimation strategies: association fitting without using instruments, MPCMR fitting where the basis function is the eigenfunctions (therefore the exposure is not transformed), and MPCMR fitting where the basis function is the polynomial functions of degree 1 (therefore with a transformed exposure). The black solid and dashed curves represent the estimated effect function results and their 95\% confidence intervals from the GMM method. The grey dashed curves represent the confidence intervals built by the Lagrange multiplier statistic. The underlying time-varying effect is the lifetime increasing effect (Scenario 3), which is represented by a blue line.
}
	\label{mf1}
\end{figure}

We first discuss the exposure-outcome model scenario 3 (the lifetime increasing effect) as a worked example. The time-varying MR results for different instrument-exposure models and estimation strategies for one single simulated dataset are shown in Figure \ref{mf1}. For the naive association approach that does not use instruments, a persistent bias is observed over time due to confounding, as the principal components are correlated with the confounders. With MPCMR, the fitted effect function curve, whether using eigenfunctions or linear polynomials as basis functions, closely resembles the true effect function. For MPCMR fitting, compared with the conventional confidence intervals based on GMM point estimates (denoted by the dashed black curves), LM-based confidence intervals (denoted by the dashed grey curves) are more conservative. The true effect function is covered by LM-based confidence intervals at all timepoints in this simulated example. MPCMR exhibits good performance in this example, irrespective of the instrument-exposure model (sinusoidal or linear). Note that MPCMR using the two different basis functions produces similar effect function estimates. This is because the eigenfunctions used (the first two eigenfunctions) are both approximately linear (see Supplementary Figure \ref{supplementary_fpca}), resulting in similar linear results for the two MPCMR strategies.

\begin{figure}[tbp]
	\centering
	\includegraphics[width=0.95\textwidth]{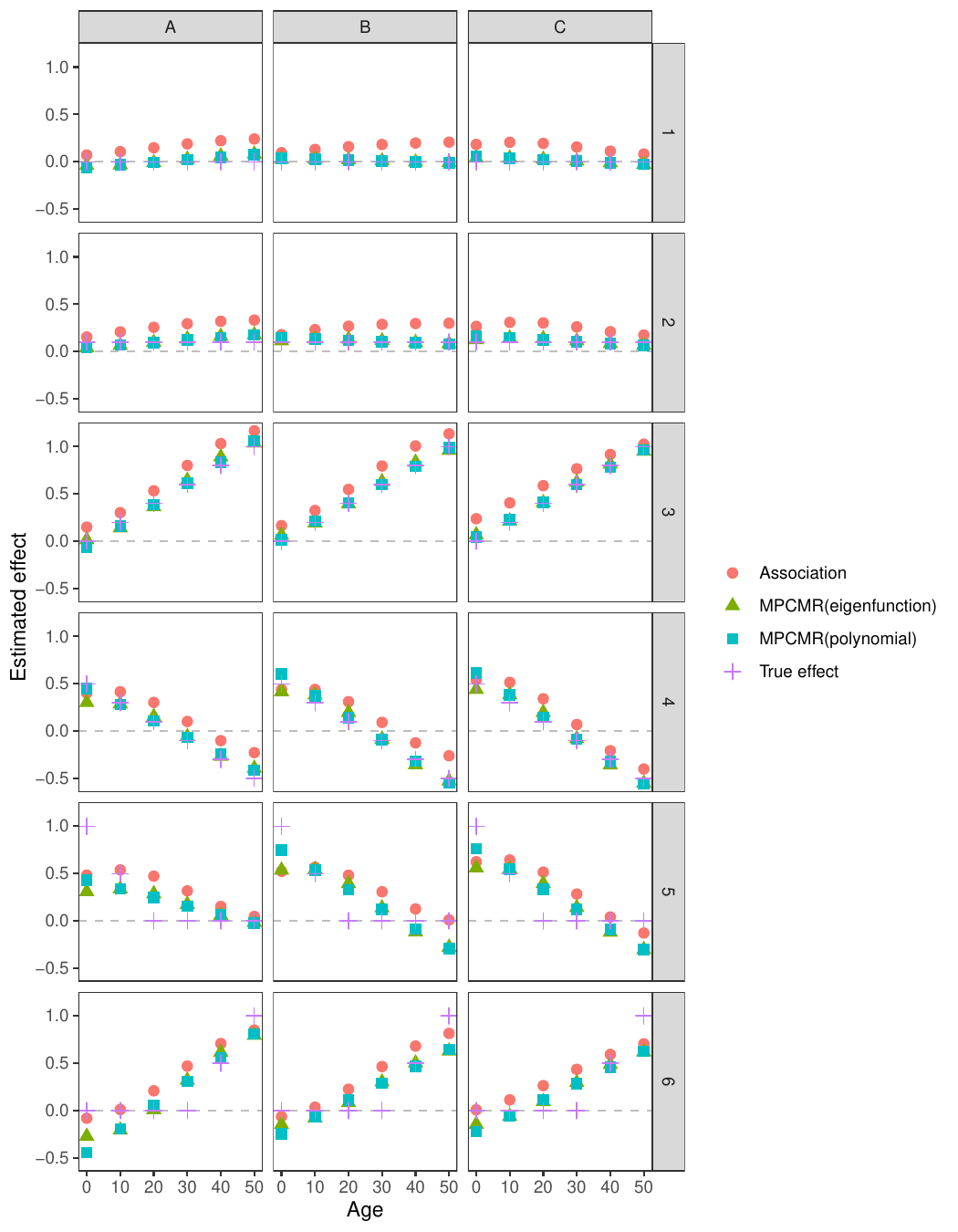}
	\caption{Mean time-varying predicted effect values using different estimation strategies under different instrument-exposure models with 1,000 simulations at 6 timepoints ($t=0,10,20,30,40,50$). The columns correspond to the instrument-exposure models A, B, and C, respectively. The rows correspond to the exposure-outcome models 1-6. Association: the association fitting without using IVs. MPCMR(eigenfunction): MPCMR fitting where the basis functions are the eigenfunctions. MPCMR(polynomial): MPCMR fitting where the basis functions are linear polynomials.
}
	\label{predict_scatterplot}
\end{figure}

Now we examine the performance of different estimating strategies in time-varying MR for all the exposure and outcome model scenarios considered. Results are provided in Figure \ref{predict_scatterplot} and Table \ref{tableres}. In all the considered timepoints, MPCMR (using eigenfunction and polynomial) shows better performance in terms of MSE, and coverage rates compared to naive estimation without using instruments (i.e. the association fitting) for outcome model scenarios 1-4, where the true effect function is linear and can be well approximated by eigenfunctions or polynomials. The coverage rate using LM statistics is at an appropriate level (approximately $0.95$) in most cases. The three exposure models (A-C) exhibit increasing instrument strength, reflected by the increasing conditional F values for each principal component. With an increase in IV strength, MPCMR usually performs better in terms of smaller MSE. When the effect function is linear, MPCMR estimation using the correct basis function (e.g., the polynomial basis) has better MSE performance than MPCMR using eigenfunctions as the basis function, which are determined by the exposure data rather than user-defined.

When the true effect function cannot be well approximated by the linear combination of basis functions (corresponding to outcome models 5-6), our effect function estimates have larger biases at some timepoints (also see Supplementary Figure \ref{supplementary_mf1}). Therefore, in addition to the IV assumptions, one should also consider the functional form of the underlying effect function and choose appropriate basis functions. Adhering to the aphorism that all models are wrong but some are useful \cite{box1976science}, the linear basis function may still offer valuable insights into determining the overall trend of time-varying effects and, consequently, the relative importance between the exposure over early and late ages on the outcome. Taking outcome scenario 5 and the left panel of Supplementary Figure \ref{supplementary_mf1} as an example, the negative slope of the fitted function supports the notion that the exposure during early ages is more likely to have a more positive effect than exposure during later ages. However, as the models are misspecified, there is considerable bias in estimation in the threshold outcome scenarios.

\begin{sidewaystable}
\centering
\scalebox{0.75}{
\begin{tabular}{c|c|c||cccc|cc||cccc|cc||cccc|cc }
  \hline
\multicolumn{3}{c|}{} & \multicolumn{6}{c|}{Scenario A} &  \multicolumn{6}{c|}{Scenario B}  & \multicolumn{6}{c}{Scenario C}  \\
\multicolumn{3}{c|}{}  & \multicolumn{6}{c|}{8.043 (PC1)  7.284 (PC2)} &  \multicolumn{6}{c|}{16.414 (PC1)  13.861 (PC2)  }  & \multicolumn{6}{c}{25.918 (PC1)  20.900 (PC2) }  \\
\multicolumn{3}{c|}{} & A1 & A2 & A3 &  \multicolumn{1}{c}{A4} & \multicolumn{1}{c}{A5} & \multicolumn{1}{c|}{A6} & B1 & B2 & B3 &  \multicolumn{1}{c}{B4} & \multicolumn{1}{c}{B5}  & \multicolumn{1}{c|}{B6}  & C1 & C2 & C3 &  \multicolumn{1}{c}{C4} & \multicolumn{1}{c}{C5}  & \multicolumn{1}{c}{C6}  \\
  \hline
   \hline
\multirow{6}{*}{$t=10$}  & \multirow{3}{*}{MSE ($10^{-2}$)}  & Association  & 1.149 & 1.166 & 1.047 & 1.361 & 0.176 & 0.022 & 1.694 & 1.722 & 1.564 & 1.990 & 0.479 & 0.140 & 4.261 & 4.325 & 4.212 & 4.648 & 2.124 & 1.313 \\
 & &  MPCMR(eigenfunction) & 0.456 & 0.468 & 0.757 & 0.374 & 3.002 & 4.630 & 0.321 & 0.345 & 0.259 & 0.928 & 0.464 & 0.857 & 0.327 & 0.364 & 0.206 & 0.931 & 0.460 & 0.551 \\
 &  &  MPCMR(polynomial) & 0.445 & 0.442 & 0.548 & 0.378 & 3.006 & 4.140 & 0.310 & 0.345 & 0.261 & 0.763 & 0.405 & 0.696 & 0.324 & 0.375 & 0.257 & 0.832 & 0.433 & 0.459 \\
\cline{2-21}
  & \multirow{3}{*}{Coverage (\%)}&  Association & 0.000 & 0.000 & 0.000 & 0.000 & 3.300 & 55.000 & 0.000 & 0.000 & 0.000 & 0.000 & 0.000 & 0.600 & 0.000 & 0.000 & 0.000 & 0.000 & 0.000 & 0.000 \\
  & & MPCMR(eigenfunction) & 98.400 & 98.400 & 97.300 & 96.800 & 70.700 & 51.400 & 98.700 & 98.500 & 96.000 & 91.500 & 97.000 & 66.800 & 98.300 & 98.200 & 95.800 & 91.000 & 97.200 & 56.600 \\
&   &  MPCMR(polynomial) & 95.300 & 95.500 & 96.100 & 94.500 & 63.200 & 48.100 & 97.100 & 97.000 & 97.100 & 89.000 & 95.400 & 69.900 & 98.000 & 97.800 & 97.600 & 90.600 & 97.000 & 62.200 \\
 \hline
 \hline
\multirow{6}{*}{$t=20$} & \multirow{3}{*}{MSE ($10^{-2}$)}  & Association & 2.192 & 2.433 & 1.764 & 4.132 & 22.229 & 4.382 & 2.543 & 2.794 & 2.166 & 4.430 & 23.160 & 5.104 & 3.765 & 4.084 & 3.485 & 5.845 & 26.465 & 6.953 \\
&  & MPCMR(eigenfunction) & 0.117 & 0.110 & 0.256 & 0.403 & 8.311 & 0.148 & 0.143 & 0.194 & 0.110 & 1.058 & 15.517 & 0.820 & 0.150 & 0.214 & 0.091 & 1.028 & 15.586 & 1.000 \\
&  & MPCMR(polynomial) & 0.067 & 0.066 & 0.093 & 0.066 & 6.147 & 0.404 & 0.092 & 0.101 & 0.070 & 0.242 & 10.947 & 1.293 & 0.104 & 0.117 & 0.072 & 0.274 & 11.277 & 1.430 \\
\cline{2-21}
& \multirow{3}{*}{Coverage (\%)} & Association  & 0.000 & 0.000 & 0.000 & 0.000 & 0.000 & 0.000 & 0.000 & 0.000 & 0.000 & 0.000 & 0.000 & 0.000 & 0.000 & 0.000 & 0.000 & 0.000 & 0.000 & 0.000 \\
&  & MPCMR(eigenfunction) & 98.100 & 97.600 & 95.000 & 76.900 & 0.200 & 96.200 & 98.800 & 98.700 & 94.600 & 60.600 & 0.000 & 74.500 & 98.500 & 98.100 & 95.500 & 59.300 & 0.000 & 62.200 \\
&  & MPCMR(polynomial) & 95.200 & 95.400 & 95.600 & 92.900 & 0.000 & 55.100 & 96.700 & 96.900 & 96.100 & 86.700 & 0.000 & 17.400 & 97.300 & 97.300 & 96.500 & 89.100 & 0.000 & 15.100 \\
 \hline
 \hline
\multirow{6}{*}{$t=30$} & \multirow{3}{*}{MSE ($10^{-2}$)} & Association & 3.576 & 3.757 & 4.032 & 4.064 & 10.025 & 22.024 & 3.325 & 3.487 & 3.795 & 3.702 & 9.437 & 21.553 & 2.432 & 2.575 & 2.727 & 2.888 & 8.046 & 18.908 \\
&  & MPCMR(eigenfunction) & 0.104 & 0.140 & 0.229 & 0.218 & 3.096 & 10.370 & 0.019 & 0.032 & 0.103 & 0.057 & 1.901 & 9.103 & 0.011 & 0.022 & 0.058 & 0.061 & 1.972 & 8.668 \\
&  & MPCMR(polynomial) & 0.081 & 0.083 & 0.053 & 0.145 & 2.465 & 9.483 & 0.014 & 0.014 & 0.011 & 0.024 & 1.501 & 8.291 & 0.007 & 0.007 & 0.006 & 0.020 & 1.506 & 8.175 \\
\cline{2-21}
&  \multirow{3}{*}{Coverage (\%)}& Association & 0.000 & 0.000 & 0.000 & 0.000 & 0.000 & 0.000 & 0.000 & 0.000 & 0.000 & 0.000 & 0.000 & 0.000 & 0.000 & 0.000 & 0.000 & 0.000 & 0.000 & 0.000 \\
&  & MPCMR(eigenfunction) & 97.800 & 97.100 & 94.000 & 88.100 & 1.000 & 0.000 & 98.900 & 97.300 & 71.800 & 83.200 & 0.000 & 0.000 & 97.900 & 95.600 & 66.600 & 70.000 & 0.000 & 0.000 \\
&  & MPCMR(polynomial) & 95.000 & 94.700 & 93.600 & 93.100 & 3.000 & 0.000 & 57.100 & 56.300 & 47.600 & 56.600 & 0.000 & 0.000 & 85.800 & 85.900 & 68.600 & 65.100 & 0.000 & 0.000 \\
 \hline
 \hline
\multirow{6}{*}{$t=40$} &\multirow{3}{*}{MSE ($10^{-2}$)} & Association & 4.928 & 4.808 & 5.370 & 3.962 & 2.300 & 4.316 & 3.924 & 3.818 & 4.277 & 3.107 & 1.597 & 3.286 & 1.275 & 1.213 & 1.391 & 0.898 & 0.199 & 0.887 \\
&  &MPCMR(eigenfunction) & 0.687 & 0.696 & 1.173 & 0.442 & 0.728 & 1.710 & 0.098 & 0.100 & 0.229 & 0.418 & 1.495 & 0.106 & 0.048 & 0.057 & 0.060 & 0.367 & 1.505 & 0.060 \\
&  & MPCMR(polynomial) & 0.489 & 0.491 & 0.426 & 0.613 & 0.730 & 0.645 & 0.077 & 0.082 & 0.085 & 0.111 & 0.813 & 0.193 & 0.035 & 0.044 & 0.058 & 0.070 & 0.833 & 0.217 \\
\cline{2-21}
& \multirow{3}{*}{Coverage (\%)} &Association & 0.000 & 0.000 & 0.000 & 0.000 & 0.000 & 0.000 & 0.000 & 0.000 & 0.000 & 0.000 & 0.000 & 0.000 & 0.000 & 0.000 & 0.000 & 0.000 & 2.800 & 0.000 \\
&  &MPCMR(eigenfunction) & 98.900 & 98.900 & 97.000 & 97.000 & 96.800 & 93.400 & 98.800 & 98.700 & 92.500 & 77.500 & 19.700 & 98.900 & 98.900 & 98.300 & 83.000 & 58.000 & 1.600 & 97.500 \\
&  &MPCMR(polynomial) & 96.400 & 96.400 & 96.500 & 96.500 & 93.200 & 94.700 & 97.300 & 97.400 & 97.500 & 95.100 & 32.600 & 90.400 & 97.900 & 98.000 & 96.800 & 95.000 & 3.300 & 65.000 \\
   \hline
    \hline
\end{tabular}
}
\caption{Time-varying MR fitting results using three estimation strategies with different combinations of the instrument-exposure models (A-C) and the exposure-outcome models (1-6). The results include the mean squared error (MSE) based on the GMM point estimates and the average coverage rate at the four timepoints ($t=10,20,30,40$) with 1,000 independent simulations. For each instrument-exposure model, the average conditional F statistics for each principal component were recorded. PC1 and PC2 represent the first and second principal components, respectively.
Association represents association fitting without using IVs. MPCMR(eigenfunction) represents MPCMR fitting where the basis functions are the eigenfunctions. MPCMR(polynomial) represents MPCMR fitting where the basis functions are linear polynomials. The coverage calculations for MPCMR are based on the identification-robust IV statistics. }
\label{tableres}
\end{sidewaystable}

\newpage
\section{Illustrative example: Systolic Blood Pressure on Urea Level}
We give an example of applying our time-varying MR method to a real data application to illustrate how our method works. Urea is a waste product removed from the blood by the kidneys. Chronic high blood pressure can impair kidney function, which may affect urea levels.

We analyzed data from UK Biobank, a population-based cohort study of UK residents ages 40-69 at recruitment \cite{sudlow2015}, who passed quality control checks as previously described \cite{astle2016allelic}. We investigate the effect of systolic blood pressure (SBP) on urea level using 258 genetic variants as instruments. These genetic variants were previously shown to be associated with a blood pressure variable in a genome-wide association study that did not include UK Biobank participants \cite{evangelou2018genetic}, thus avoiding bias due to winner's curse \cite{taylor2014mendelian}. 

We first investigate the effect of SBP on urea level in a standard univariable MR  model ignoring time, where both SBP and urea level are treated as time-invariant. The SBP value measured in mmHg at recruitment is used as the exposure, and urea measured in mmol/L at recruitment is used as the outcome. Univariable analyses are conducted in 349,883  participants, with blood pressure and urea measured at study baseline. Supplementary Figure \ref{UVMR_scatterplot_urea} shows the scatter plot of the genetic associations with the exposure and the outcome using the 258 SNPs. The univariable MR effect estimate is $0.008$ (95\% CI: $0.004, 0.012$) mmol/L per $1$ mmHg higher genetically-predicted SBP. This suggests that higher genetically-predicted SBP levels are associated with higher urea levels. The estimate may be interpreted as a cumulative effect, under the assumption that the genetic associations with the exposure are approximately constant \cite{labrecque2019interpretation}. However, a positive cumulative effect cannot provide insight on any time-varying effects. Therefore, we use the MPCMR method to investigate the possible effect pattern over time.

For the time-varying analysis, we use SBP measurements from primary care records on 158,310 individuals, and their urea measurements taken at baseline. To ensure there was sufficient data to obtain precise estimates at all timepoints, we considered blood pressure measurements from age 50 to 75 years. Participants without a SBP measurement in this time window were dropped from analyses. To reduce computational complexity, we rounded all measurements to the nearest tenth of a year.

The data used for MPCMR fitting contains 101,279 individuals. We calculated the eigenfunctions and individual principal components from FPCA, using the package \emph{fdapace}. The eigenfunction information is given in Supplementary Figure \ref{Eigenfunctions}. The first four principal components have a cumulative fraction of variance explained larger than $95\%$, with the first, second, third, and fourth components explaining $65.9 \%$, $ 18.5 \% $, $ 8.2\% $, and $ 4.7\% $ of the variance in exposure measurements. We use the first two eigenfunctions and the linear polynomials as the basis functions in MPCMR fitting to mitigate bias from weak instruments. 

\begin{figure}[htp]  
	\centering
	\includegraphics[width=1.00\textwidth]{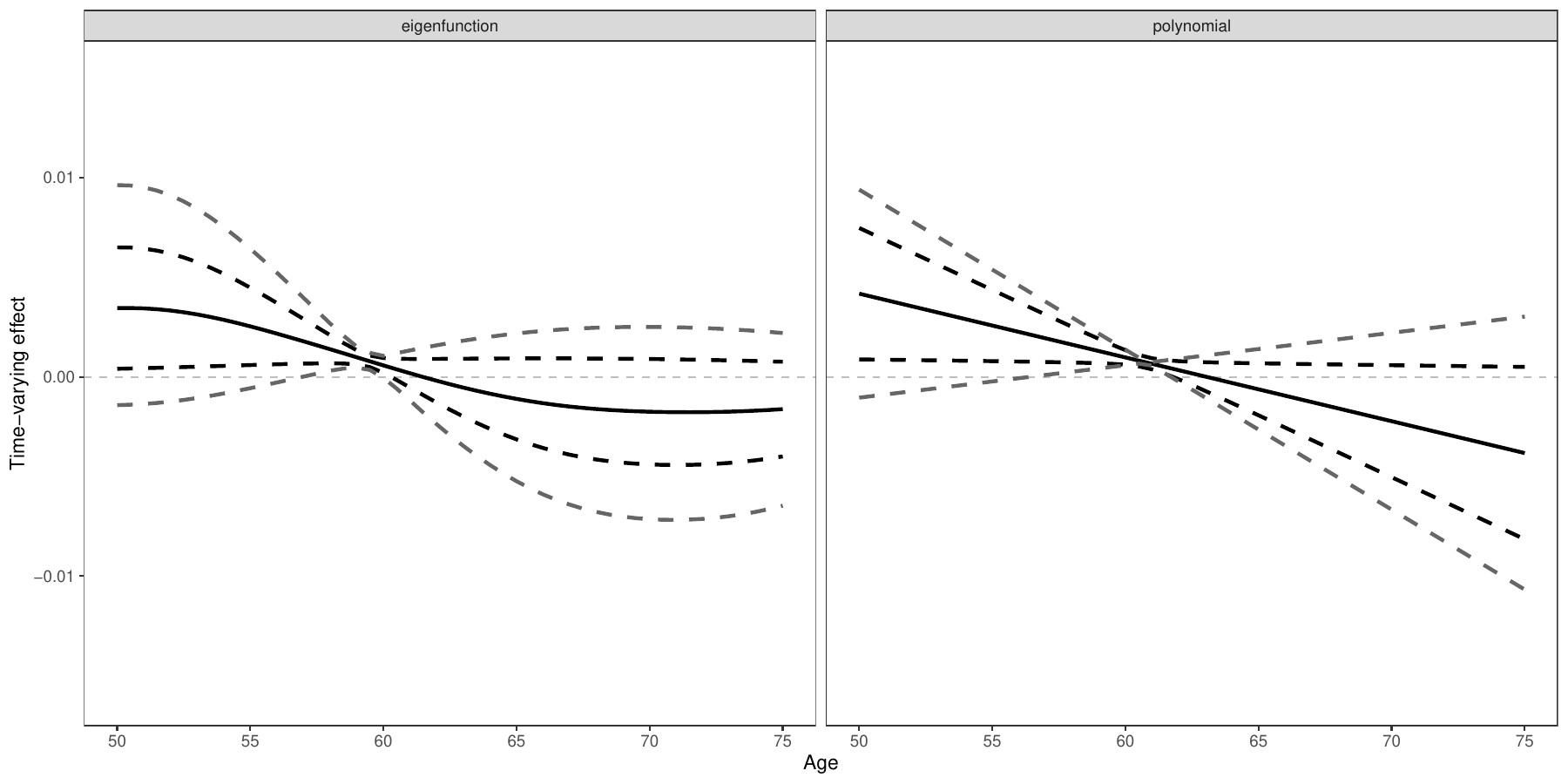} 
	\caption{Application results using the first two eigenfunctions (left panel) and the linear polynomials (right panel) as the basis functions. The black curve represents the fitted time-varying effect from the MPCMR method, while the dashed lines represent 95\% confidence intervals derived from GMM (black) and Lagrange multipliers (grey). The effect unit is mmol/L per $1$ mmHg higher genetically-predicted SBP.}
	\label{MPCMR_fit_real}
\end{figure}

The functional fitting results are given in Figure \ref{MPCMR_fit_real}. The estimated effect (represented by the solid black curve) is decreasing over time and significant and positive over some time points. Both eigenfunction and linear polynomials fitting with the GMM 95\% confidence interval (the black dashed curves) support that the early-age effect of SBP is positive and significant, while the later-age effect is not significant.
This indicates that SBP at the beginning of the time window is more likely to have a more positive effect than SBP towards the end of the time window. 
However, the results still indicate much uncertainty, and the estimated effect is significant for only a short period around 60 years as indicated by the identification-robust 95\% confidence intervals (the grey dashed curves). The high uncertainty is due to the weak-instrument problem as there is a strong pattern of linear dependence in the genetic associations with the first and second principal components. The conditional F statistics are $9.46$ and $1.52$ for the two principal components. Overall, we cannot made any strong conclusions about the time-varying nature of the effect of SBP on urea levels.

\section{Discussion}\label{Discussion}
In this paper, we have introduced a novel method for Mendelian randomization to study the time-varying effects of a modifiable exposure on the outcome. Our method is based on continuous-time modelling with a functional effect objective, so it should not suffer from violation of the IV exclusion restriction assumption caused by the model misspecification arising from using exposure measurements at limited measured timepoints. A continuous-time model has previously been proposed for time-varying MR studies \cite{cao2016mendelian,labrecque2019interpretation} but there was no feasible estimation method proposed. We have proposed a general framework for dealing with such problem relying on functional dimensional reduction ideas.
Our method initially constructs a multivariable MR model using functional principal component analysis summarizing time-varying exposure information. We then back out time-varying effect estimation and inference based on the MVMR model estimates. We considered weak-instrument-robust methods so that we can obtain reliable results regardless of instrument strength. We demonstrated with a variety of simulation scenarios that our method can accurately estimate time-varying effect functions, given that the correct form of the basis function is used.

To the best of our knowledge, all the current methods proposed in time-varying MR studies are time-discrete. For example, a previous paper considered effect estimation using multiple exposure values, but it only estimated a constant causal effect \cite{cao2016mendelian}. Other past papers  assumed the effect function can be correctly expressed as a constant effect at limited timepoints \cite{shi2021mendelian,sanderson2022estimation}. All of those additional assumptions can lead to misspecification problems and yield misleading MR results \cite{labrecque2019interpretation,tianestimation}. By contrast, we treat the time-varying effect function as the objective of interest since time is continuous, and so our approach can provides a more general way for conducting time-varying analyses.

In order to estimate the effect function, we perform dimension reduction to transform the exposure trajectories (or equivalently the genetic associations with the exposure) into nonfunctional variables. We employed FPCA, a data-driven method that borrows information across individuals and avoids making strong parametric model assumptions for the exposure trajectory or instrument-exposure associations. Our approach allows for sparse and different exposure measurement time points for each individual, which is common in practice. We adopted identification-robust inference and have demonstrated that valid conclusions can be obtained by our method for linear time-varying functions. It is important to consider weak instrument-robust inference in time-varying MR analyses, since it may be difficult to find genetic predictors of the exposure at multiple time points with appropriate linear independence.

Although our MPCMR approach allows for a flexible parametric form to learn the effect function, we have presented simulation scenarios using eigenfunction and linear basis functions, which are adequate for estimating linear effect functions, but are unable to estimate more complex functions. This choice is based on practical considerations as FPCA may only return a relatively small number of principal components explaining the exposure trajectory. Increasing the number of principal components used for MPCMR can worsen the weak-instrument problem, as the added principal components explain a lower proportion of exposure variation and so could be less associated with the instruments. This presents a trade-off in MPCMR estimation: if one desires a more flexible estimate of the time-varying function, they may suffer more from weak instrument issues. While if one uses fewer PCs to enable model fitting with strong IVs, this requires use of a simpler form for fitting the effect function. In particular, if we only have two principal components, then we can only estimate two parameter exposure--outcome models. This means that estimating complex time-varying functions is likely to be infeasible in practice. With the increasing availability of detailed time-stamped diagnoses and longitudinal data from electronic health records in large biobanks, such as UK Biobank \cite{bycroft2018uk} and FinnGen (\url{https://www.finngen.fi/en}), we believe time-varying MR analysis has the potential to become more widely implemented in future studies. However, this will require the existence of genetic variants having effects on exposure trajectories that vary in time. This is a natural limitation to the feasibility of any approach for time-varying MR studies.

There are some methodological limitations in this study.
First, to extract reliable principal component information, FPCA requires individual exposure values measured at multiple time points, with each time point containing a sufficient number of samples. This means the data should be longitudinal over the time region of interest. 
Second, FPCA may not explain a sufficient proportion of the variation in the exposure with the desired number of principal components, and the remaining ignored principal components could induce structural pleiotropy, which may be a concern for instrument validity. In our example, the first two principal components explained around 85\% of the variability in the exposure trajectories, which is good from the perspective of avoiding structural pleiotropy, but only allows the estimation of simple exposure--outcome functions.
Third, as in any application of MR, some instruments may have direct effects on the outcome and may invalidate our estimation, which requires the IV assumptions to hold. Additionally, using a truncated, clipped, or incomplete time region may violate the IV assumptions, even if all IVs are valid in the original complete time region. 
Fourth, uncertainty in estimating the eigenfunctions is ignored in our approach. Uncertainty can be reduced by increasing the sample size or the number of measured time points. Additionally, we assume that the same eigenfunctions hold for all individuals.
Finally, although our method aims to be generally applicable in time-varying MR studies, it still requires some assumptions to ensure the model is correct. Future methodological work should consider the relaxation of these assumptions.

In summary, we have presented a novel method with continuous-time modelling that combines data-driven approaches and weak IV robust methods to obtain reliable estimators for simple time-varying MR effects. Our method offers a general framework for studying the effects of time-varying modifiable exposures, but is limited by the availability of IVs with time-varying effects on the exposure. Since most exposures of interest are time-varying, our method has the potential to be an important element for future MR analyses. However, there are intrinsic limitations on the level of detail for time-varying effects that can learned in practice.

\section*{Acknowledgements}
For the purpose of open access, the authors have applied a Creative Commons Attribution (CC BY) licence to any Author Accepted Manuscript version arising from this submission.

\section*{Code availability}
The R package to apply the methods proposed in this manuscript is available at \url{https://github.com/HDTian/TVMR}, along with R code that reproduces the empirical results in this manuscript.

\newpage
\thispagestyle{empty}

\singlespacing
\bibliographystyle{plain}
\bibliography{All_TVMR}

\clearpage
\doublespacing
\thispagestyle{empty}
\renewcommand{\thesection}{A\arabic{section}}
\renewcommand{\thesubsection}{A.\arabic{subsection}}
\renewcommand{\thetable}{A\arabic{table}}
\renewcommand{\thefigure}{A\arabic{figure}}
\renewcommand{\theequation}{S\arabic{equation}}
\setcounter{table}{0}
\setcounter{figure}{0}
\setcounter{equation}{0}
\renewcommand{\tablename}{Supplementary Table}
\renewcommand{\figurename}{Supplementary Figure}
\setcounter{section}{0}
\setcounter{subsection}{0}

\newpage
\section*{Supplementary Figures}
\begin{figure}[H]
	\begin{minipage}{0.33\linewidth}
		\centering
		\includegraphics[height=5.0cm]{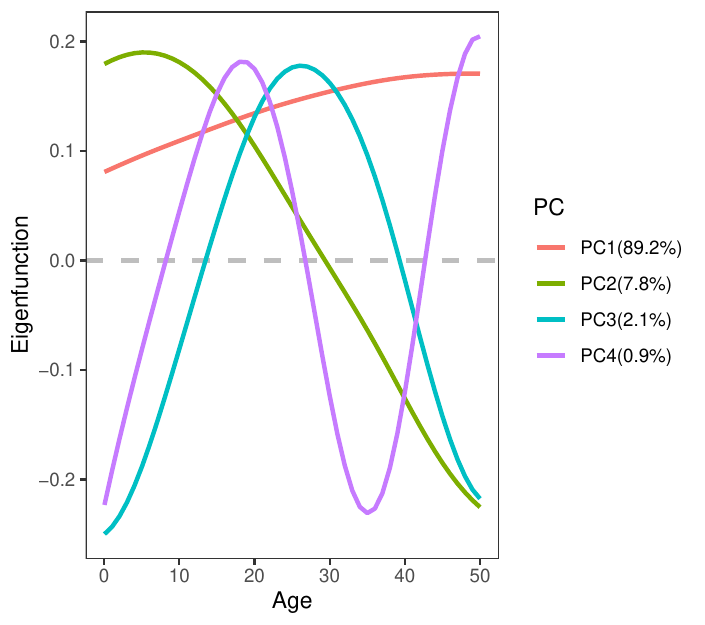} 		
	\end{minipage}
	\begin{minipage}{0.33\linewidth}
		\centering
		\includegraphics[height=5.0cm]{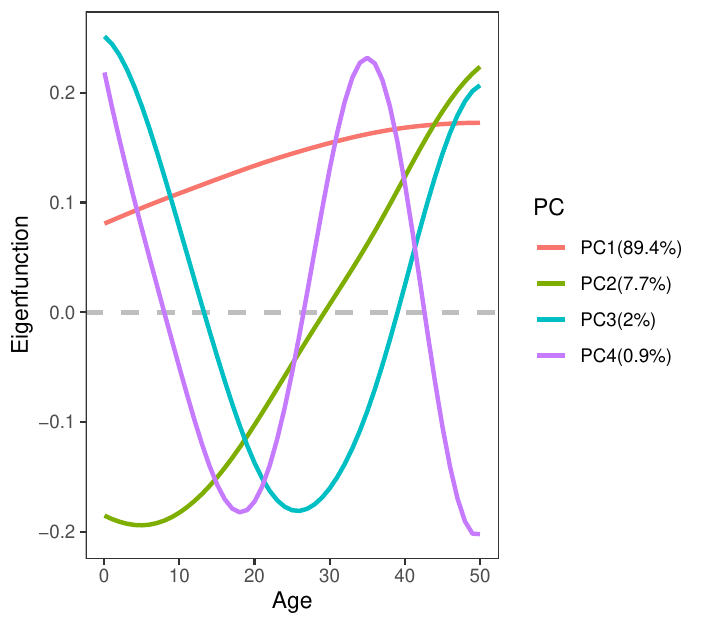} 	
	\end{minipage}       
 \begin{minipage}{0.33\linewidth}
		\centering
		\includegraphics[height=5.0cm]{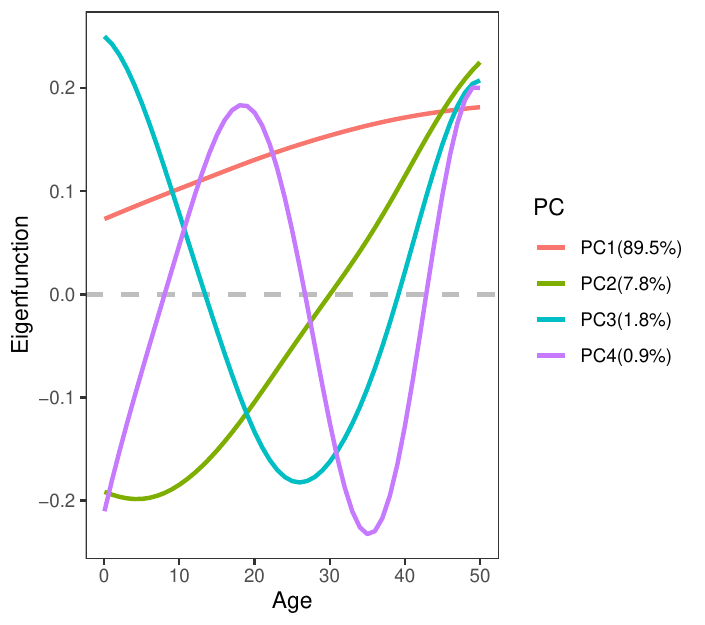} 	
	\end{minipage}
	\caption{The eigenfunction results of functional principal components in one simulation example under the three exposure model scenarios A-C. The left, middle and right panel corresponds to the instrument-exposure scenarios A, B and C, respectively. In all cases, the first and second eigenfunctions, denoted by the red and green curves, are approximately linear and cumulatively explain more than $95\%$ of the variations.}
	\label{supplementary_fpca}
\end{figure}

\begin{figure}[H]
	\begin{minipage}{0.5\linewidth}
		\centering
		\includegraphics[height=7.5cm]{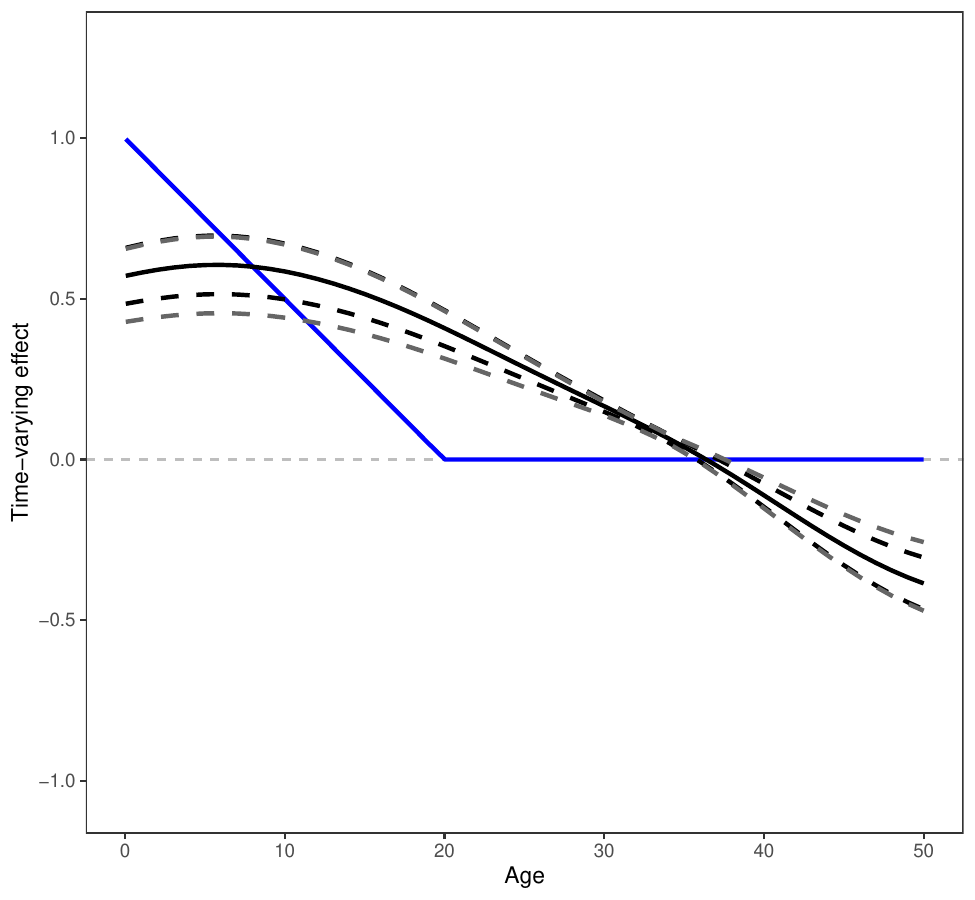} 		
	\end{minipage}
	\begin{minipage}{0.5\linewidth}
		\centering
		\includegraphics[height=7.5cm]{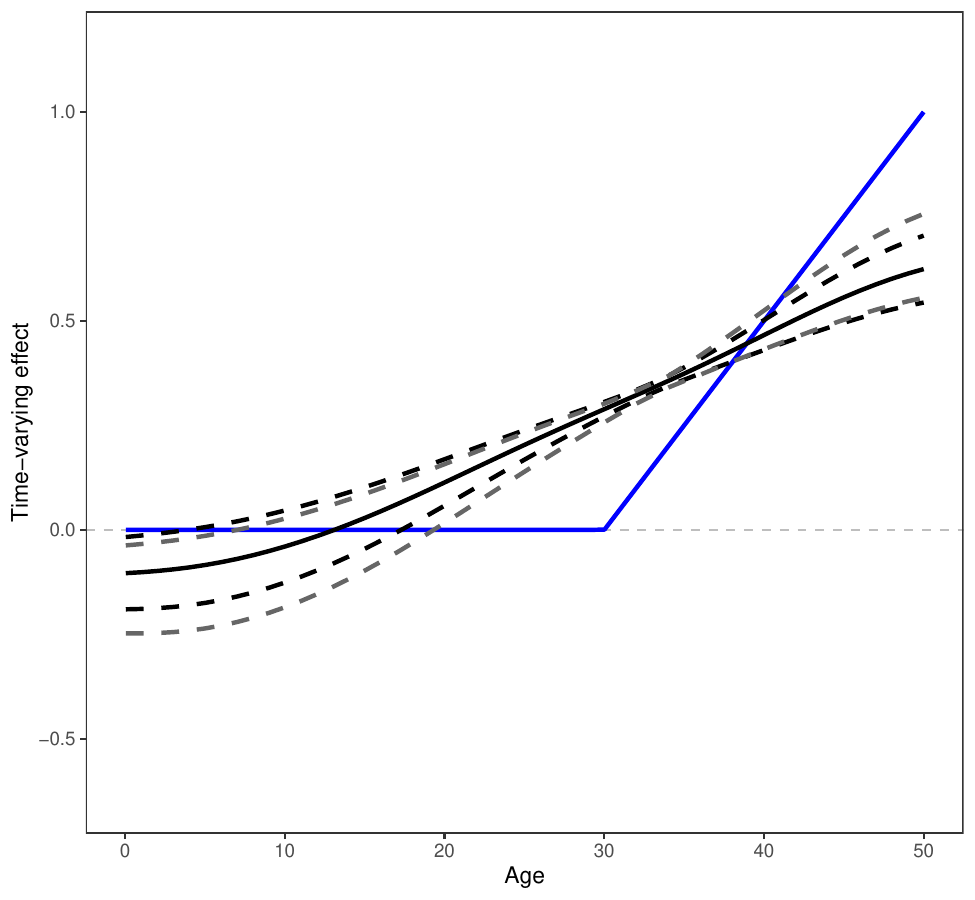} 	
	\end{minipage}       
	\caption{The time-varying MR fitting results using eigenfunctions as the basis functions under the exposure model scenario C. Left: the outcome model scenario 5. Right: the outcome model scenario 6. The true effect functions, which are expressed by the blue curves, cannot be represented by the eigenfunctions in an additive linear way, thereby causing bias in most timepoints.}
	\label{supplementary_mf1}
\end{figure}

\begin{figure}[H]  
	\centering
	\includegraphics[width=0.65\textwidth]{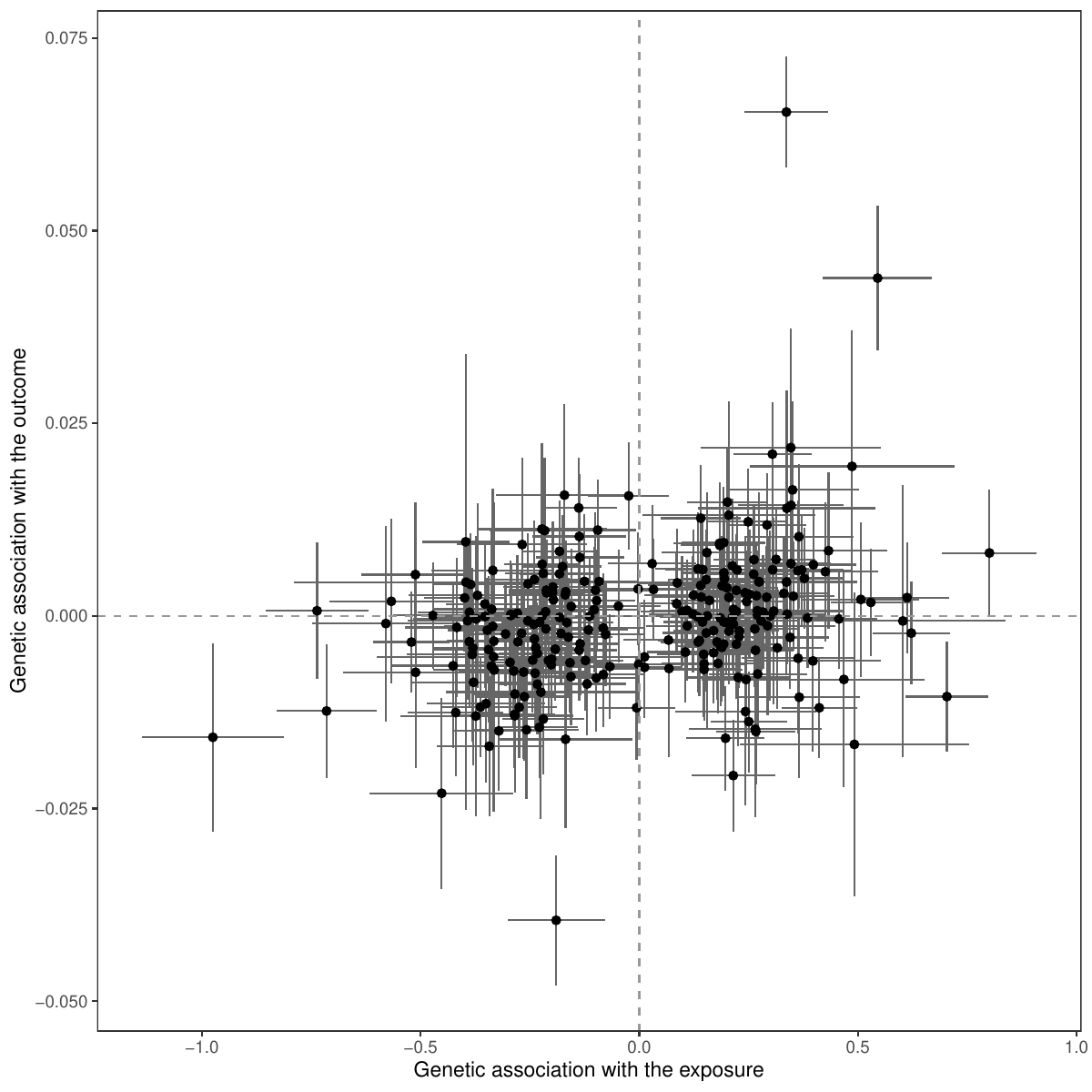} 
	\caption{Scatter plot of the estimated genetic association with the exposure and outcome for 258 SNPs in the univariable Mendelian randomization fitting.}
	\label{UVMR_scatterplot_urea}
\end{figure}

\begin{figure}[H]  
	\centering
	\includegraphics[width=0.6\textwidth]{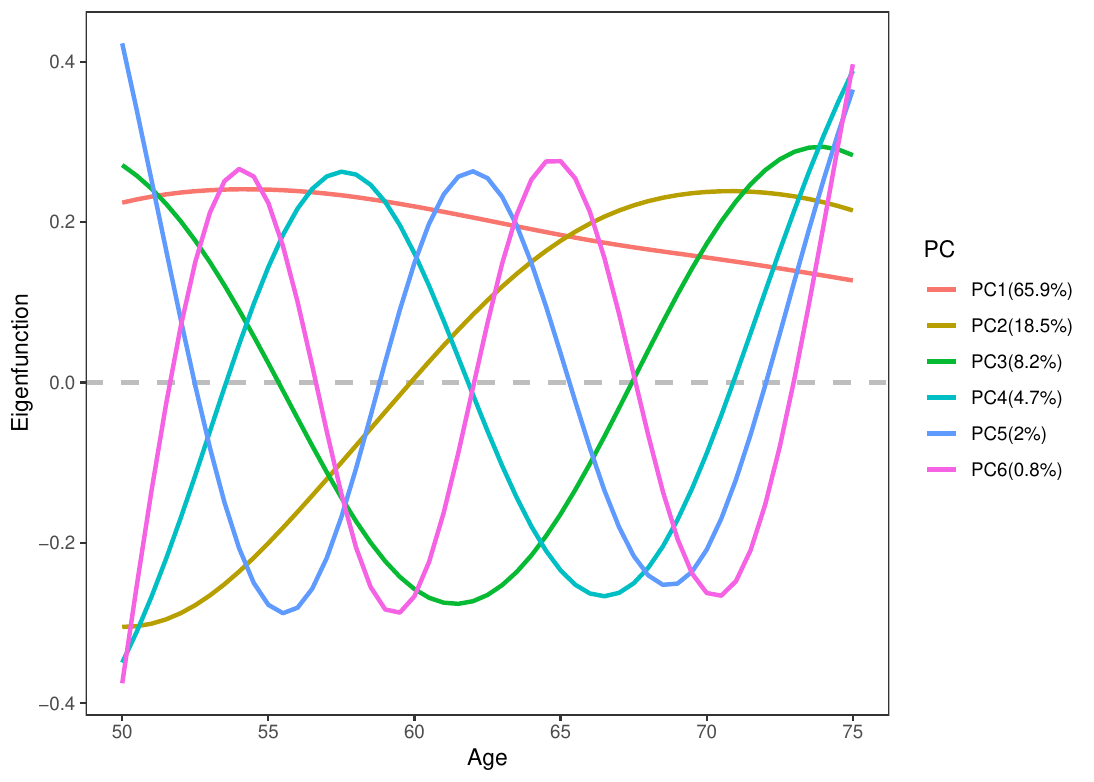} 
	\caption{The eigenfunctions of FPCA for the SBP data over the age region $[50,75]$. PC stands for the principal component, followed by the corresponding fraction variance explained.}
	\label{Eigenfunctions}
\end{figure}

\newpage
\section*{Supplementary Texts}
\subsection*{Text S1: Basic identification equation}
We illustrate the key idea of identifying the time-varying effect, which is a functional objective, using instrumental variable (IV) methods. Assuming the IV core assumptions are valid, and the structural outcome is
\begin{equation}
	Y_T=	\int_{0}^{T} \beta_T(t) X(t)dt+g_{Y,T}(U, \epsilon_Y)
\end{equation}
which can be regarded as an infinitely dimensional multivariable Mendelian randomization (MVMR) model, where the multiple exposures represent the time-varying exposures at different time points. $U$ contains both the time-invariant and time-varying confounders up to the time point $T$ (e.g. $ \{ U(t); 0 \leq t\leq T \} $).
Let's assume there are $J$ independent genetic variants denoted as ${G_j; j=1,\ldots,J}$. The inverse variance weighted (IVW) method involves two-stage fittings. In the second stage, we perform a regression of the outcome on the instruments (with possible covariates for adjustment). The estimated parameter for the $j$-th instrument follows an asymptotic distribution (here, we ignore the uncertainty associated with the standard errors):
\begin{equation}
	\hat{\theta}_j \overset{D}{\to} \mathcal{N}(  \theta_j , s.e.(\hat{\theta}_j)^2 )
\end{equation}
where $ \theta_j = cov( Y_T, G_j )/var(G_j) =cov(	\int_{0}^{T} \beta_T(t) X(t)dt, G_j )/var(G_j) = 	\int_{0}^{T} \beta_T(t)\alpha_j(t) dt $ and $ \alpha_j(t):= cov(G_j,X(t))/var(G_j) $. It is easy to know $ cov( \hat{\theta}_{j_1}, \hat{\theta}_{j_2} )=0  $ for any $j_1\neq j_2$. Therefore, we have the following equation that enables the identification of $ \beta_T(t) $: $ \hat{\theta}_j  \sim  \mathcal{N}( \int_0^T \beta_T(t) \alpha_j(t) dt,s.e.(\hat{\theta}_j)^2)
$,
which is the ideal fitting equation for IVW results. However, since $\alpha_j(t)$ is unknown, it needs to be estimated from the data. This estimation can be obtained through the first-stage IVW fitting, which involves regressing $X(t)$ on the instruments (with possible covariates for adjustment). If the uncertainty associated with the estimated functional objective $\hat{\alpha}_j(t)$ is small, we can proceed with the following fitting regression
\begin{equation}\
\hat{\theta}_j = \int_0^T \beta_T(t) \hat{\alpha}_j(t)dt + \epsilon_j \qquad \epsilon_j\sim \mathcal{N}(0, s.e.(\hat{\theta}_j)^2)\qquad j=1,\ldots,J
\end{equation}
which is the standard fitting equation for obtaining IVW results. Note that this equation represents an ill-posed inverse problem, which means that $\beta_T(t)$ may not be consistently estimated even if we have a consistent estimator of $\theta_j$ and $\alpha_j(t)$ for $j=1,\ldots,J$. Therefore, we cannot directly express the final estimated term (the effect function, $\hat{\beta}_T(t)$, in time-varying MR) in a manner similar to univariable MR or MVMR without making additional assumptions about $\beta_T(t)$.

\subsection*{Text S2: Functional principal component analysis}
We briefly introduce the functional principal component analysis (FPCA) and its basic properties. For more details, see the review paper \citeappendix{wang2016functional}. Assume the individual function curve $X_i(t), i=1,\ldots,n$ is a smooth and square-integrable function, where $X_i(t)$ is a functional objective like the exposure trajectory. The FPCA treats $X_i(t)$ as a realization of the random function $ X(t) $ over the region of interest $[0,T]$. Denote the mean exposure function by $ \mathbb{E}(X(t))= \mu(t) $. Define the covariance function at any two timepoints $(s,t) $ as $ cov( X(s), X(t) ) $. Assume that the covariance function can be expressed by the orthogonal expansion in the $L^2$ sense
\begin{equation} \label{G}
cov( X(s),X(t)  ) = \sum_{k=1}^{\infty}  \lambda_k \phi_k(s) \phi_k(t)
\end{equation}
where $ \lambda_1\geqslant\lambda_2\geqslant\ldots \geqslant \lambda_{\infty}  \geqslant 0$ are called eigenvalues and $   \phi_1(\cdot),\phi_2(\cdot), \ldots, \phi_{\infty}(\cdot) $ are the orthonormal eigenfunctions satisfying in sequence
$$
\phi_1 =  \argmax_{   \|  \phi \|  =1     }\left\lbrace    var\left(   \int_{0}^T ( X(t)-\mu(t) ) \phi(t)dt     \right)  \right\rbrace
$$
$$  \vdots $$
$$
\phi_k =  \argmax_{   \|  \phi \|  =1; < \phi,\phi_j >=0; j=1,\ldots,k-1     }\left\lbrace    var\left(   \int_{0}^T ( X(t)-\mu(t) ) \phi(t)dt     \right)  \right\rbrace
$$
$$  \vdots $$
where $ \|  \phi \| := \left[    \int_{0}^T \phi(t)^2 dt      \right] ^{\frac{1}{2}}   $ and $ < \phi,\phi_j > :=  \int_{0}^T \phi(t) \phi_j(t)   dt     $. The $k$-th eigenfunction is the function maximizing the variance of the projected value of the exposure curve over this function across individuals and is also orthogonal to the previous eigenfunctions. We define the principal component associated with the $k$-th eigenfunction as 
$
\xi_k =  \int_{0}^T ( X(t)-\mu(t) ) \phi_k(t)dt
$. For any $k_1$-th and $k_2$-th principal components, it is easy to know
\begin{equation}
\begin{split}
cov(\xi_{k_1} , \xi_{k_2}  ) =& cov(  \int_{0}^T X(t) \phi_{k_1}(t)dt, \int_{0}^T X(t) \phi_{k_2}(t)dt     )
=  cov(  \boldsymbol{  \phi }_{k_1}^T \boldsymbol{  X }\delta , \boldsymbol{  \phi }_{k_2}^T \boldsymbol{  X }\delta  )
= \delta \boldsymbol{  \phi }_{k_1}^T    cov( \boldsymbol{  X }, \boldsymbol{  X } )  \boldsymbol{  \phi }_{k_2} \delta \\
\overset{\left\langle 1\right\rangle}{=}&  \delta \boldsymbol{  \phi }_{k_1}^T    \left( \sum_{k=1}^{\infty} \lambda_k  \boldsymbol{  \phi }_{k} \boldsymbol{  \phi }_{k}^T   \right)  \boldsymbol{  \phi }_{k_2} \delta
=  \sum_{k=1}^\infty \lambda_k( \boldsymbol{  \phi }^T_{k_1}\boldsymbol{  \phi }_{k}   \delta )( \boldsymbol{  \phi }^T_{k_2}\boldsymbol{  \phi }_{k}   \delta )\\
=& \sum_{k=1}^\infty \lambda_k \int_0^T \phi_{k_1}(t) \phi_k(t)dt \int_0^T \phi_{k_2}(t) \phi_k(t)dt =0  \qquad \textrm{for } k_1 \neq k_2
\end{split}
\end{equation}
due to the eigenfunction properties, where $ \boldsymbol{  \phi }_k = ( \phi_k(\delta) \,\phi_k(2\delta) \, \ldots \,  \phi_k(T)   )^T$, $ \boldsymbol{  X } = ( X(\delta) \,X(2\delta) \, \ldots \,  X(T)   )^T$ and $\delta$ is a small time grid for numerical approximation. Equation $\left\langle 1\right\rangle $ is due to the orthogonal expansion \eqref{G}.

The variance of principal components in the $k$-th eigenfunction direction is the $k$-th eigenvalues. That is, $   var( \xi_k )=\lambda_k  $. According to \eqref{G} and the orthonormal eigenfunction property, we can know
\begin{equation}\label{equ}
\int_{0}^T 	cov( X(s),X(t)  ) \phi_k(s) ds = \lambda_k \phi_k(t)
\end{equation}
The curve can be then expressed by the Karhunen-Lo\'eve theorem as
$
X(t)- \mu(t)  =  \sum_{k=1}^{\infty}  \xi_k \phi_k(t)
$.
Hence, we can approximately express each individual curve with the first $M$ eigenfunctions explaining enough variation (e.g., $>95\%$)
\begin{equation} \label{approximation}
X_i(t) \approx \mu(t) + \sum_{m=1}^{M}\xi_{m,i} \phi_m(t)
\end{equation}
where $ \xi_{m,i} $ is the principal component value of the $i$-th individual w.r.t the $m$-th eigenfunction.

\subsection*{Text S3: Principal components analysis through conditional expectation}
We briefly introduce how to draw FPCA with sparse measured samples. One can see the original paper \citeappendix{yao2005functional} proposing PACE in detail. Assume we have the individual data where the individual measured exposure are $ \{  X_{ij} ; j=1,\ldots,N_i  \} $ where $  X_{ij} $ represents the $j$-th observation at timepoint $ t_{ij}$ of the $i$-th individual. PACE makes the following assumptions:
\begin{itemize}
	\item[(i)] The measured exposure could contain the measurement error for the accurate exposure but it is in an additive independent way; that is, $ X_{ij}= X_i( t_{ij} )+ \epsilon_{ij} $ and $ \epsilon_{ij} \overset{\textrm{i.i.d}}{\sim}\mathcal{N}(0, \sigma^2 )  $.
	\item[(ii)] The individual exposure is a random function expressed as the FPCA (via the Karhunen-Lo\'eve theorem): $ X_i(t_{ij})=\mu( t_{ij} )  +  \sum_{k=1}^{\infty} \xi_{k,i}\phi_k( t_{ij} ) $ where the distribution of the principal components $\{  \xi_k ; k=1,2,\ldots \}$ are assumed to be Gaussian.
	\item[(iii)] The individual measured timepoints $ \{  t_{ij}; j=1,\ldots,N_i \} $ and the number of the measured timepoint, $N_i$, are independent of the measurement errors and the principal components.
\end{itemize}
Therefore, the individual random variables $  \{ \boldsymbol{  X }_i ,  \boldsymbol{  \xi }_i  \} $, where $  \boldsymbol{  X }_i=( X_{i1} \, \cdots \, X_{iN_i} )^T $ and $ \boldsymbol{  \xi }_i=( \xi_{1,i} \, \cdots \, \xi_{\infty,i} )^T   $, follow the multivariate normal distribution:
\begin{equation}
\left( \begin{array}{c}
\boldsymbol{  \xi }_i  \\
\boldsymbol{  X }_i
\end{array} \right)
 \sim
\mathcal{N}(
\left( \begin{array}{c}
 \boldsymbol{  0 } \\
 \boldsymbol{  \mu }_i
\end{array} \right)
,
\left( \begin{array}{cc}
var( \boldsymbol{  \xi }_i  ) &\boldsymbol{  \Gamma } \\
\boldsymbol{  \Gamma }^T &   \boldsymbol{  \Sigma }_{\boldsymbol{  X }_i}
\end{array} \right)   )
\end{equation}
where $\boldsymbol{  \Gamma }_{k,j}:=cov(  \xi_{k,i}  , X_i(t_{ij}) )  = cov( \xi_{k,i}, \xi_{k,i}\phi_k(t_ij)  )  =\lambda_k  \phi_k(t_{ij}) $ due to the FPCA property.  $\lambda_k=var( \xi_k ) $ is the eigenvalue for the $k$-th principal component, $\boldsymbol{  \Sigma }_{\boldsymbol{  X }_i} = var(\boldsymbol{  X }_i  )  $, and $ \boldsymbol{  \mu }_i = (   \mu(t_{i1}) \, \cdots \, \mu(t_{iN_i}) )^T $. The conditional expectation of the $k$-th principal components given the other observations for an individual is therefore
\begin{equation}
\mathbb{E}(   \xi_{k,i} | \boldsymbol{  X }_i  )  = \lambda_k \boldsymbol{  \phi }^T_{k,i}  \boldsymbol{  \Sigma }_{\boldsymbol{  X }_i}^{-1} ( \boldsymbol{  X }_i - \boldsymbol{  \mu }_i  )
\end{equation}
where $ \boldsymbol{  \phi }_{k,i}= (   \phi_k(t_{i1})  \, \cdots \, \phi_k(t_{iN_i}) )^T $. In practice, we first draw the curve and surface smoothing with observed population data to obtain the estimates $ \hat{\mu}(t) $ and $  \widehat{cov}(X(t),X(s) ) $ (as well as $ \hat{\sigma} $ and $ \hat{\boldsymbol{  \Sigma  }}_{\boldsymbol{  X }} $); then derive the estimated eigenfunctions $ \{ \hat{\phi}_k(t) \} $ and eigenvalues $\{ \hat{\lambda}_k \}$ based on the equation \eqref{equ}; finally, the conditional expectation of principal components with the plug-in estimates is
\begin{equation}
\hat{\xi}_{k,i}=\hat{\mathbb{E}}(   \xi_{k,i} | \boldsymbol{  X }_i  )  = \hat{\lambda}_k \boldsymbol{  \hat{\phi} }^T_{k,i}  \boldsymbol{  \hat{\Sigma} }_{\boldsymbol{  X }_i}^{-1} ( \boldsymbol{  X }_i - \hat{\boldsymbol{  \mu }}_i  )
\end{equation}
which is used as the $k$-th principal component score of the $i$-th individual. Similar to before, we can choose the first $K$ principal components for reasonably good approximation and name $K$ as the number of the primary principal components. $K$ can be chosen according to the fraction-of-variance-explained (FVE), cross-validation style, AIC-type criteria, etc.

\subsection*{Text S4: Two-sample summary-level data}
Since we transform the time-varying analysis problem into MVMR problems, one advantage of this approach is that it allows for a two-sample and summary-level data setting. This means that the exposure can come from one dataset, while the outcome can come from another dataset \citeappendix{burgess2015using,davies2018reading}. The more flexible data structure significantly increases the applicability of our methods for time-varying studies. In most cases, public longitudinal data do not measure both the exposure and outcome together, as the exposures and outcomes are often defined after the study protocol. In contrast, the most widely recognized models for time-varying studies, namely the structural (nested) mean models \citeappendix{robins1989analysis,robins1994correcting,robins2000marginal}, typically require one-sample data for fitting. The corresponding IV models \citeappendix{michael2023instrumental} also require complete longitudinal information for the instrument, exposure, and outcome.

Our MPCMR models can also be implemented with summary-level data. For summary statistics, we additionally assume that the instrument-exposure model is (where the principal components are either the original PC or the transformed ones with transformation matrix, depending on the choice of basis functions):
\begin{equation}
    {\xi}_k = {\alpha}_0 + \boldsymbol{\alpha}_k^T \boldsymbol{Z} + V_k \quad k=1,2,\ldots,K
\end{equation}
with $ \mathbb{E}( V_{k1}V_{k2}| \boldsymbol{Z} ) $ does not depend on $ \boldsymbol{Z} $ for any $k_1$ and $k_2$, which is called the homoskedasticity assumption. Therefore, we have the instrument-outcome model:
$
    Y = {\theta}_0 + \boldsymbol{\theta}^T \boldsymbol{Z} + U \, , \, k=1,2,\ldots,K
$
with $  \mathbb{E}(  U |\boldsymbol{Z}    ) = 0 $ and $ \mathbb{E}(   U^2 | \boldsymbol{Z} ) $ does not depend on $ \boldsymbol{Z} $. We can construct the moment functions with summary statistics
$
   \hat{\boldsymbol{g}}(  \boldsymbol{\beta}) =  \hat{   \boldsymbol{\theta} } - \hat{ \boldsymbol{\alpha} }\boldsymbol{\beta}
$
where the estimated association $ \hat{ \boldsymbol{\theta} } \in \mathbb{R}^{ J \times 1 } $ and $ \hat{ \boldsymbol{\alpha} } \in \mathbb{R}^{ J \times K } $ can be from either the univariable regression or multivariable regression. Then the effect function $ \beta(t)$ can be estimated with the similar GMM procedures introduced in the main text.

\subsection*{Text S5: MPCMR estimation with IVW}
We illustrate the MPCMR estimation with the classical IVW framework. We first assume the effect shape satisfies the parametric form: $  \beta(t) = \sum_{l=1}^L \gamma_l b_l(t) $ where $  \{ \beta_l\} $ are known basis functions and $ L\leq K $ for identification.

Review that the structural equation of MPCMR \eqref{MVMR} to form the MVMR respective to the parameters of interest $ \{ \gamma_l\}$
\begin{equation}
\begin{split}
Y_i =& \beta_0+ \int_{0}^{T} \beta(t) X_i(t)dt + g_Y(U_i,\epsilon_{Y,i}) \\
=&   \beta_0 + \sum_{l=1}^L \gamma_l \int_0^T b_l(t) X_i(t) dt + g_Y(U_i,\epsilon_{Y,i})
\end{split}
\end{equation}
where the MVMR estimates $ \hat{\gamma}_l $ is based on the IVW regression, where we assume the SNPs are pruned to be or assumed to be not correlated with each other,
\begin{equation}
\begin{split}
\hat{\theta}_j =&\sum_{l=1}^L \gamma_l \frac{cov( G_j , \int_0^T b_l(t) X(t) dt  )}{var(G_j)}  + \epsilon_j  \qquad \epsilon_j  \sim \mathcal{N}(0, s.e.(\hat{\theta}_j )^2 \tau^2) \\
=& \sum_{l=1}^L \gamma_l \frac{cov( G_j , \int_0^T b_l(t) \left[\mu( t )  +  \sum_{k=1}^{K} \xi_{k}\phi_k( t )  \right]  dt  )}{var(G_j)}  + \epsilon_j \\
=&  \sum_{l=1}^L \gamma_l \sum_{k=1}^{K} \frac{    cov( G_j ,      \xi_{k}    ) }{var(G_j)}\int_0^T b_l(t)\phi_k( t )  dt  + \epsilon_j
\end{split}
\end{equation}
with the plug-in estimates $\frac{    \widehat{cov}( G_j ,      \xi_{k}    ) }{\widehat{var}(G_j)}  $ from the regressing the PCs on the SNPs, and the standard numeric estimates $ \int_0^T b_l(t)\phi_k( t )  dt  \approx  \sum_i^I \phi_k(t_i )b_l(t_i)T/I $ for the uniform grids $\{t_i\}$. The IVW regression is therefore
\begin{equation}
\hat{\boldsymbol{  \theta }} =  \hat{ \boldsymbol{  \alpha } } \boldsymbol{  B }   \boldsymbol{  \gamma }  + \boldsymbol{  \epsilon }   \qquad \boldsymbol{  \epsilon } \sim \mathcal{N}( \boldsymbol{  0 } ,\tau^2 \boldsymbol{  \Sigma }  )
\end{equation}
where  $  \boldsymbol{  B }  = \int_{0}^{T}  \boldsymbol{  \phi }(t) \boldsymbol{  b }^T(t) dt  \in \mathbb{R}^{K\times L} $, $\hat{ \boldsymbol{  \alpha } } \in \mathbb{R}^{J\times K} $ is the instrument-PCs association estimates and $ \boldsymbol{  \Sigma } $ is the diagonal matrix with $ \boldsymbol{  \Sigma }_{j,j} = s.e.( \hat{\theta}_j )^2 $. The estimators are
\begin{equation}\label{naiveEst}
\hat{  \boldsymbol{  \gamma } } = (  \boldsymbol{  B }^T\hat{ \boldsymbol{  \alpha } }^T    \boldsymbol{  \Sigma }^{-1}  \hat{ \boldsymbol{  \alpha } } \boldsymbol{  B }    )^{-1} \boldsymbol{  B }^T\hat{ \boldsymbol{  \alpha } }^T \boldsymbol{  \Sigma }^{-1} \hat{\boldsymbol{  \theta }}
\end{equation}
with the estimator variance matrix $  (  \boldsymbol{  B }^T\hat{ \boldsymbol{  \alpha } }^T    \boldsymbol{  \Sigma }^{-1}  \hat{ \boldsymbol{  \alpha } } \boldsymbol{  B }    )^{-1} \hat{\tau}^2  $ and the pointwise functional error $ \boldsymbol{  b }^T(t)  (  \boldsymbol{  B }^T\hat{ \boldsymbol{  \alpha } }^T    \boldsymbol{  \Sigma }^{-1}  \hat{ \boldsymbol{  \alpha } } \boldsymbol{  B }    )^{-1} \hat{\tau}^2  \boldsymbol{ b }(t)  $. There are typically two common choices for the parametric form of $b_l(t)$:
\begin{enumerate}
	\item[1.] The polynomial basis function:  $   b_l(t)=t^{l-1} $, $l=1,2,\ldots$
	\item[2.] The eigenfunction system: $ b_l = \phi_l(t) $, $l=1,2,\ldots$
\end{enumerate}

One special kind of parametric assumption used in recent TVMR studies is to assume the constant-effect form that $ \beta(t)= \gamma $ for any timepoint \citeappendix{cao2016mendelian}, which leads to a one-parameter problem and greatly simplifies the estimation model. In our context, such an assumption means $  \boldsymbol{  B }  = \int_{0}^{T}  \boldsymbol{  \phi }(t)   1  dt =  \boldsymbol{  1 }  \in \mathbb{R}^{K}  $ as the eigenfunction property. The IVW regression becomes the UVMR regression model: $   \boldsymbol{  \theta }_j = \left(  \sum_k^K \hat{  \alpha  }_{j,k}   \right)    \gamma   +   \epsilon_j  $ where $ \epsilon_j \sim \mathcal{N}(0, s.e.( \theta_j )^2 ) $.

\subsection*{Text S6: Time-varying IV validity and basis function testing}
In Mendelian randomization, some genetic variants may be invalid. This can be caused by the possible genetic correlation with confounders (hence the violation of exchangeability), or the genetic pleiotropy phenomenon \citeappendix{davey2003mendelian} (hence the violation of the exclusion restriction). One may wish to assess the IV validity. The IV validity test is also called the over-identification test \citeappendix{hansen1982large}, assessing whether all instruments are estimating the same causal effect parameters ($\{ \beta^\ast_k\}$ or $ \{ \gamma_l \}$). One special source making an instrument invalid is the basis function assumptions for $\beta(t)$. When using the incorrect basis function in $\boldsymbol{B}$, the instrument for the transformed principal components will fail to estimate the same parameter and therefore be invalid. When some instrument is invalid, the over-identification should be rejected. Note that we cannot test the validity when all instruments used are invalid in the same way, therefore caution should be given that the IV validity test assesses only the coherency rather than validity \citeappendix{parente2012cautionary}. However, due to the ignorable possibility of all instruments being invalid in the same way, we can regard the over-identification test as the IV validity test.

We can test the IV validity using many methods, including Cochran's Q test \citeappendix{cochran1954combination}, Sargan J test \citeappendix{sargan1988testing}, or other goodness-of-fit tests. We take the Q test as an example due to its popular usage for summary-level MR analysis. Assume the null hypothesis $H_0$: all instruments are valid and $ \beta(t) =  \sum_{l=1}^L \gamma_l b_l(t) $ is the correct shape form.
We assume that each single instrument is uncorrelated with each other. This independent instrument assumption can be guaranteed in practice by pruning the SNPs to near independence (e.g. by using the genetics analysis software PLINK \citeappendix{purcell2007plink}) or compressing them to principal components by ordinary PCA \citeappendix{batool2022disentangling}. When independent instruments cannot be guaranteed, one can choose the Sargan J test instead (this is also embedded in our package TVMR). Let the number of instruments and the (untransformed) PCs be $J$ and $K$ (for identification, we require $J\geq K \geq L$). The corresponding test statistic is:
\begin{equation} \label{Qstatistic}
Q = \sum_{j=1}^{J} \frac{(  \hat{\theta}_j - \hat{\boldsymbol{  \alpha } }_j^T \boldsymbol{  B } \hat{  \boldsymbol{  \gamma } } )^2}{  s.e.( \hat{\theta}_j )^2 + \hat{ \boldsymbol{  \gamma } }^T \boldsymbol{  B }^T  \boldsymbol{  \Sigma }_{\alpha,j} \boldsymbol{  B } \hat{  \boldsymbol{  \gamma } } - 2 \hat{ \boldsymbol{  \gamma } }^T \boldsymbol{  B }^T   \boldsymbol{  \Gamma }_j   }
\end{equation}
where $\hat{\boldsymbol{ \alpha } }_j $ is the estimated instrument-PCs associations for the $j$-th instrument, and $ \boldsymbol{ \Sigma }_{\alpha,j} $ is the corresponding variance matrix. $ \boldsymbol{ \Gamma }_j = \text{cov}( \hat{ \boldsymbol{ \alpha } }_j, \hat{\theta}_j ) $. Under the null hypothesis $H_0$ that the instruments are valid and the basis function assumption is correctly specified, the statistic follows a $\chi^2_{ J- L} $ distribution asymptotically. The details for calculating each element in this statistic can be seen in Supplementary Text S9. Particularly, the quantities $ \hat{\boldsymbol{ \gamma } } $ can be estimated by either the maximal profile likelihood procedure \citeappendix{zhaoMRRAPS,wang2021causal} or the iterative steps (given in Supplementary Text S9) that are commonly used in summary-data MR studies \citeappendix{bowden2019improving}.

Note that one can use the statistic \eqref{Qstatistic} to test the IV validity only, regardless of the basis function assumptions. This is because, no matter the form of $\beta(t)$, the instruments should estimate the same parameters $\{ \beta^\ast_k  \}$ under the null hypothesis. This is equivalent to the classical MVMR model, in which case the Q statistic equals the formula \eqref{Qstatistic} by setting $ \boldsymbol{B}=\boldsymbol{I}$ or any other transformation matrix $\boldsymbol{B} \in \mathbb{R}^{K\times K}$ that is full rank.

\subsection*{Text S7: Time-varying IV strength testing}
Another critical problem in Mendelian randomization and instrumental variable analysis is the weak instrument. The weak instrument refers to the instrument's strength being insufficient to support meaningful and reliable results. A weak instrument can lead to ill-identification problems and make the inference on the parameters of interest invalid and unreliable \citeappendix{andrews2019weak}. When fitting univariable MR models, the strength of the instrument depends on the strength and uncertainty of the genetic associations, which can be expressed by its effect size or the F statistics \citeappendix[ch.~16]{hernan2020causal}. When fitting multivariable MR models, the concept of a weak instrument needs to take into account the possible linear dependence that the genetic association with a single exposure (PC in our context) may be linearly expressed by the genetic association with the remaining exposures (PCs). This leads to larger estimation error and, therefore, weak instrument bias. Such linear dependence can be expressed by either conditional F statistics \citeappendix{sanderson2016weak, sanderson2019examination} or the Q statistic \citeappendix{sanderson2021testing}. The same test statistic \eqref{Qstatistic} can be used for testing the instrument strength.

When using the Q statistic to test the instrument strength for the $k$-th PC with the remaining PCs, we treat the $k$-th PC as the 'outcome,' and the remaining PCs are the 'exposures.' We then let $ \hat{\theta}_j  = \hat{\boldsymbol{  \alpha }}_{j,k} $ and use the updated $ \hat{\boldsymbol{  \alpha }}_j  \leftarrow \hat{\boldsymbol{  \alpha }}_{j,-k} $ with the corresponding adjusted standard errors. Set $ \boldsymbol{  B }=\boldsymbol{  I } \in \mathbb{R}^{(K-1)\times(K-1)}$, and the testing degree of freedom is $J- (K-1)$. Strong evidence for rejection indicates strong instrument strength.

The IV strength and validity test is provided by our package TVMR.

\subsection*{Text S8: Related proof of LM statistic}

We show that $ \hat{\Delta}_k=\hat{ \mathbb{E} }_n(  - \boldsymbol{Z} \xi_k  (Y - \boldsymbol{\xi}^T \boldsymbol{\beta}_0) \boldsymbol{Z}^T  ) $ is the consistent estimator of
$ \Delta_k := cov(
 \sqrt{n} \hat{ \boldsymbol{g} }( \boldsymbol{\beta}_0 ) , \sqrt{n} \hat{\boldsymbol G}_k )  $.
 \begin{proof}
 Without loss of generality, we let all variables be centered; hence
\begin{equation}
    \begin{split}
      \Delta_k  =   & cov(
 \sqrt{n} \hat{ \boldsymbol{g} }( \boldsymbol{\beta}_0 ) , \sqrt{n} \hat{\boldsymbol G}_k )  \\
        = & n \,cov(   \hat{\mathbb{E}}_n(  \boldsymbol{Z} (  Y - \boldsymbol{\xi}^T \boldsymbol{\beta}_0 ) ),  \hat{\mathbb{E}}_n(    - \boldsymbol{Z} \xi_k    )   ) \\
        =&  cov(     \boldsymbol{Z}( Y- \boldsymbol{\xi}^T \boldsymbol{\beta}_0) , - \boldsymbol{Z} \xi_k   )  \quad \textrm{due to i.i.d samples} \\
        =& \mathbb{E}(  - \boldsymbol{Z} \xi_k  (Y - \boldsymbol{\xi}^T \boldsymbol{\beta}_0) \boldsymbol{Z}^T  )
    \end{split}
\end{equation}
 \end{proof}

We show that $   cov( \sqrt{n} \hat{D}_k( \boldsymbol{\beta}_0 ) , \sqrt{n}\hat{\boldsymbol{g} }( \boldsymbol{\beta}_0 ) ) \overset{p}{\to} \boldsymbol{0}$.
\begin{proof}
review that $  \hat{D}_k( \boldsymbol{\beta}_0 ) = \hat{\boldsymbol{G}}_k - \hat{\Delta}_k(  \boldsymbol{\beta}_0 )^T \hat{\Omega}(\boldsymbol{\beta}_0)^{-1} \hat{\boldsymbol{g}}( \boldsymbol{\beta}_0 ) $; hence
\begin{equation}
    \begin{split}
     cov( \sqrt{n} \hat{D}_k( \boldsymbol{\beta}_0 ) , \sqrt{n}\hat{\boldsymbol{g} }( \boldsymbol{\beta}_0 ) ) =& cov( \sqrt{n} \hat{\boldsymbol{G}}_k  , \sqrt{n}\hat{\boldsymbol{g} }( \boldsymbol{\beta}_0 ) )  - cov( \sqrt{n}  \hat{\Delta}_k(  \boldsymbol{\beta}_0 )^T \hat{\Omega}(\boldsymbol{\beta}_0)^{-1} \hat{\boldsymbol{g}}( \boldsymbol{\beta}_0 ) ,\sqrt{n}\hat{\boldsymbol{g} }( \boldsymbol{\beta}_0 )   )\\
      \overset{p}{\to}&   cov( \sqrt{n} \hat{\boldsymbol{G}}_k  , \sqrt{n}\hat{\boldsymbol{g} }( \boldsymbol{\beta}_0 ) )  -   {\Delta}_k(  \boldsymbol{\beta}_0 )^T {\Omega}(\boldsymbol{\beta}_0)^{-1}  cov( \sqrt{n}\hat{\boldsymbol{g}}( \boldsymbol{\beta}_0 ) ,\sqrt{n}\hat{\boldsymbol{g} }( \boldsymbol{\beta}_0 )   ) \\
      =& \Delta_k( \boldsymbol{\beta}_0 )^T-  {\Delta}_k(  \boldsymbol{\beta}_0 )^T {\Omega}(\boldsymbol{\beta}_0)^{-1}{\Omega}(\boldsymbol{\beta}_0)  \quad \textrm{by definition}\\
      =&  \boldsymbol{0}
    \end{split}
\end{equation}
\end{proof}

\subsection*{Text S9: Q statistic testing details}
We provide the testing inference details regarding the parametric MPCMR models. For UVMR, a similar inference procedure can be found in the papers \citeappendix{zhaoMRRAPS,bowden2019improving}; for MVMR, see the papers \citeappendix{wang2021causal,sanderson2021testing}. Assume the null hypothesis $H_0$ is: $ \beta(t) =  \sum_{l=1}^L \gamma_l b_l(t) $ for fixed basis functions $ \{ b_l(t) \}$ and parameters $ \{ \gamma \}$ to be estimated. We assume that the SNPs are pruned to be uncorrelated to each other; therefore, the covariance of the genetic associations from different SNPs to any phenotypes (PCs, outcome, whether same or different) is neglectably zero. We can build the joint likelihood for the estimated associations with the phenotypes for each SNP
\begin{equation}
\left( \begin{array}{c}
\hat{\theta}_j  \\
\hat{\boldsymbol{  \alpha }}_j
\end{array} \right)     \sim \mathcal{N}(
\left( \begin{array}{c}
\boldsymbol{  \alpha }_j^T  \boldsymbol{  B }   \boldsymbol{  \gamma }  \\
\boldsymbol{  \alpha }_j
\end{array} \right)  ,
\left( \begin{array}{cc}
s.e.( \hat{\theta}_j )^2 &  \boldsymbol{\Gamma}_j^T  \\
\boldsymbol{\Gamma}_j & \boldsymbol{\Sigma}_{\alpha,j}  \\
\end{array} \right)		)   \qquad j=1,2,\ldots,J
\end{equation}
Where $ \boldsymbol{\Sigma}_{\alpha,j}   $ is the variance matrix for $ \hat{\boldsymbol{  \alpha }}_j$. We can approximate that $ \boldsymbol{\Sigma}_{\alpha,j}   $ is a diagonal matrix with the $k$-th element equal to $  s.e.(  \hat{\alpha}_{j,k} )^2 $. $ \boldsymbol{  \Gamma }_j := cov( \hat{ \boldsymbol{  \alpha } }_j, \hat{\theta}_j ) $. When the exposure and the outcome come from two samples, $ \boldsymbol{  \Gamma }_j  = \boldsymbol{  0 }$; otherwise $\boldsymbol{  \Gamma }_j$ is estimated by the approximation under the overlapping-samples setting
\begin{equation}
\boldsymbol{  \Gamma }_{j,k} = cov(  \hat{\alpha}_{j,k} ,  \hat{\theta}_j ) = \frac{n_s}{n_1 n_2} \frac{1}{var(G_j)} cov( v^{PC}_k , v^{Y} )
\end{equation}
with the relevant plug-in estimates, where $ v^{PC}_k$ and $ v^{Y} $ are the residuals of the regression on the genetic variants from the $k$-th PC and the outcome, respectively, under the common dataset; $n_1$ and $n_2$ are the sample size of the PC data and the outcome data, $ n_s $ is the shared sample size.
\begin{proof}
	Assume the genetic variants are uncorrelated with each other. Consider the regression for the two phenotypes $ \xi_k $ and $Y$ with the respective (possibly overlapping) samples:
	\begin{equation}
	\xi_{k,i}  =  \alpha_0 +  \boldsymbol{  \alpha }^T \boldsymbol{  G }_i +  v_{1,i}    \qquad i=1,\ldots,n_1
	\end{equation}
	\begin{equation}
	Y_i  =  \theta_0  +  \boldsymbol{  \theta }^T \boldsymbol{  G }_i +  v_{2,i}  \qquad i=1,\ldots,n_2
	\end{equation}
	where $ v_1 \overset{\textrm{def}}{=} v^{PC}_k $ and $  v_2 \overset{\textrm{def}}{=} v^{Y}  $; $ \alpha_j = cov( G_j, \xi_k )/var(G_j) $ and $\theta_j = cov( G_j, Y )/var(G_j)$. When $\boldsymbol{  G }$ are binary (like in most MR cases), it is easy to know $  \mathbb{E}( G^p v_1 ) = 0 $ and $  \mathbb{E}( G^p v_2 ) = 0 $ for any order $p=0,1,\ldots$. Therefore, the genetic association estimators for the $j$-th genetic variant are
	\begin{equation}
	\hat{  \alpha }_{j,k} = \frac{\widehat{cov}( G_j , \xi_k )}{  \widehat{var}( G_k ) } =
	\frac{  (\overline{  G_j \xi_k })_{n_1} - (\overline{ G_j})_{n_1} (\overline{ \xi_k })_{n_1} }{  \widehat{var}( G_k ) }
	\end{equation}
	\begin{equation}
	\hat{  \theta }_{j} = \frac{\widehat{cov}( G_j , Y )}{  \widehat{var}( G_k ) }=
	\frac{  (\overline{  G_j Y })_{n_2} - (\overline{ G_j})_{n_2} (\overline{ Y })_{n_2} }{  \widehat{var}( G_k ) }
	\end{equation}
	The approximation has the following steps
	\begin{equation}
	\begin{split}
	cov(	\hat{  \alpha }_{j,k} , \hat{  \theta }_{j}   )  \overset{\left\langle 1\right\rangle }{=}
	& \frac{1}{var(G_j )^2}  cov[  (\overline{  G_j \xi_k })_{n_1} - (\overline{ G_j})_{n_1} (\overline{ \xi_k })_{n_1},    (\overline{  G_j Y })_{n_1} - (\overline{ G_j})_{n_1} (\overline{ Y })_{n_1}   ]   \\
	=&  \frac{1}{var(G_j )^2}  cov[   \frac{1}{n_1} \sum_{i=1}^{n_1} G_{j,i}( \xi_{k,i}^C  + (\overline{ \xi_k })_{n_1} )  -  (\overline{ G_j})_{n_1} (\overline{ \xi_k })_{n_1}  ,
	\frac{1}{n_2} \sum_{i=1}^{n_1} G_{j,i}( Y_i^C  + (\overline{ Y })_{n_2} )  -  (\overline{ G_j})_{n_2} (\overline{ Y })_{n_2}    ] \\
	\overset{\left\langle 2 \right\rangle }{=} &  \frac{1}{var(G_j )^2}  cov[ \frac{1}{n_1} \sum_{i=1}^{n_1} G_{j,i} \xi_{k,i}^C ,    \frac{1}{n_2} \sum_{i=1}^{n_2} G_{j,i} Y_{i}^C   ] \\
	\overset{\left\langle 3 \right\rangle }{=}&   \frac{n_s}{n_1 n_2}\frac{1}{var(G_j )^2}  cov(  G_j \xi_k^C , G_j Y^C  )  \\
	\overset{\left\langle 4 \right\rangle }{=} & \frac{n_s}{n_1 n_2}\frac{1}{var(G_j )^2}  cov(  \alpha_{j,k} G^2_j  + G_j v_1  , \theta_j G^2_j  + G_j v_2  ) \\
	=&    \frac{n_s}{n_1 n_2}\frac{1}{var(G_j )^2}  \left(    \alpha_{j,k} \theta_j var( G_j^2 )+cov(  G_jv_1,G_jv_2 ) \right) \\
	\overset{\left\langle 5 \right\rangle }{\approx} &  \frac{n_s}{n_1 n_2}\frac{1}{var(G_j )^2} cov(  G_jv_1,G_jv_2 )\\
	\overset{\left\langle 6 \right\rangle }{=} &  \frac{n_s}{n_1 n_2}\frac{1}{var(G_j )} cov(  v_1,v_2 )
	\end{split}
	\end{equation}
	Where the centred phenotypes $ \xi_k^C$ and $Y^C $ in Equation $\left\langle 2 \right\rangle $ corresponds to the equations $  \xi_k - \mathbb{E}( \xi_k ) =  \boldsymbol{  \alpha }^T \boldsymbol{  G } +  v_{1} $ and $  Y - \mathbb{E}( Y ) =  \boldsymbol{  \theta }^T \boldsymbol{  G } +  v_{2}   $, and $ n_s$ in Equation $\left\langle 3 \right\rangle $ is the shared sample size. In Equation $\left\langle 4 \right\rangle $, we assume $  \mathbb{E}( G^2_j G_{j_\ast} v_1  ) = 0  $ and $  \mathbb{E}( G^2_j G_{j_\ast} v_2  ) = 0  $ for any $j_\ast $. We approximate $  \alpha_{j,k} \theta_j \approx 0   $ in Equation  $\left\langle 5 \right\rangle $, as the genetic associations in Mendelian randomization are usually small. In equation $\left\langle 6 \right\rangle $, we assume $ \mathbb{E}( G_j^2 v_1 v_2) = \mathbb{E}(G_j^2  ) \mathbb{E}(v_1v_2) $, which equals to $  var( G_j ) cov(   v_1 ,v_2) $.
\end{proof}
Hence, the estimator for $ \boldsymbol{  \Gamma }_{j,k} $ is
\begin{equation}
\hat{\boldsymbol{  \Gamma }}_{j,k} = \widehat{cov}(  \hat{\alpha}_{j,k} ,  \hat{\theta}_j ) = \frac{n_s}{n_1 n_2} \frac{1}{\widehat{var}(G_j)} \widehat{cov}( v^{PC}_k , v^{Y} ) \approx \frac{1}{n_1 n_2} \frac{1}{\widehat{var}(G_j)}  \sum_{i=1}^{n_s} ( \xi_{k,i} - \hat{\xi}_{k,i} )( Y_i  -\hat{Y}_i )
\qquad k=1,\ldots,K
\end{equation}
where $ \widehat{cov}( v^{PC}_k , v^{Y} )$ is estimated by the share sample part, and we can approximate that $ \frac{1}{n_s} \sum_{i=1}^{n_s} ( \xi_{k,i} - \hat{\xi}_{k,i} )  \sum_{i=1}^{n_s}( Y_i  -\hat{Y}_i ) \approx 0 $. The statistic is therefore
\begin{equation}
Q = \sum_{j=1}^{J} \frac{(  \hat{\theta}_j - \hat{\boldsymbol{  \alpha } }_j^T \boldsymbol{  B } \hat{  \boldsymbol{  \gamma } } )^2}{  s.e.( \hat{\theta}_j )^2 + \hat{ \boldsymbol{  \gamma } }^T \boldsymbol{  B }^T  \boldsymbol{  \Sigma }_{\alpha,j} \boldsymbol{  B } \hat{  \boldsymbol{  \gamma } } - 2 \hat{ \boldsymbol{  \gamma } }^T \boldsymbol{  B }^T   \boldsymbol{  \Gamma }_j   }
\end{equation}
Under the null that the parametric model is correctly specified, the statistic follows the chi-squared distribution $ \chi^2_{ J- L} $. The estimator of $\hat{   \boldsymbol{  \gamma } } $ based on the statistic is regarded as the robust estimator. The iterative steps for deriving the robust estimator and Q statistic is
\begin{enumerate}
	\item[(1)] Obtain the initial estimates $\hat{  \boldsymbol{  \gamma } }_{(0)}$ based on Equation \eqref{naiveEst}, which is to minimize
	\begin{equation}
	Q(\boldsymbol{  \gamma };\boldsymbol{  \gamma }_\ast)= \sum_{j=1}^{J} \frac{(  \hat{\theta}_j - \hat{\boldsymbol{  \alpha } }_j^T \boldsymbol{  B }   \boldsymbol{  \gamma }  )^2}{  s.e.( \hat{\theta}_j )^2 +  \boldsymbol{  \gamma }_\ast ^T \boldsymbol{  B }^T  \boldsymbol{  \Sigma }_{\alpha,j} \boldsymbol{  B }  \boldsymbol{  \gamma }_\ast  - 2 \boldsymbol{  \gamma }_\ast^T \boldsymbol{  B }^T   \boldsymbol{  \Gamma }   }
	\end{equation}
	with $ \boldsymbol{  \gamma }_\ast  = \boldsymbol{  0 } $ (this is equivalent to the standard MVMR fitting).
	
	\item[(2)] Minimize $ Q(   \boldsymbol{  \gamma }; \hat{\boldsymbol{  \gamma }}_{(0)} ) $ to obtain $ \hat{\boldsymbol{  \gamma }}_{(1)} $
	
	\item[(3)] Repeat the above step and update the iterative estimates $  \hat{ \boldsymbol{  \gamma } }_{(n)} $ until the estimates are stable.
\end{enumerate}
The corresponding $Q$ statistic value with the plug-in MPCMR robust estimates is then used to evaluate the parametric shape assumptions or other testing objectives (like IV strength and validity as introduced in the previous supplementary texts).

\singlespacing
\bibliographystyleappendix{plain}
\bibliographyappendix{All_TVMR}

\end{document}